\documentclass[%
 aps,
 amsmath,amssymb,reprint,
 pra,superscriptaddress]{revtex4-2}
\newcommand\trick[1]{}
\usepackage{graphicx}
\usepackage{dcolumn}
\usepackage{bm}

\usepackage{siunitx}
\usepackage[normalem]{ulem}
\DeclareSIUnit\gauss{G}
\DeclareSIUnit{\au}{a.u.}

\usepackage[usenames,dvipsnames]{xcolor}
\usepackage[colorlinks=true,urlcolor=blue,citecolor=magenta,linkcolor=magenta]{hyperref}
\usepackage[final]{microtype}
\usepackage{xcolor}
\def\red#1{\textcolor{black}{#1}}

\def\blaa#1{{#1}}

\graphicspath{{./},{gfx/}}

\newcommand{\ket}[1]{\ensuremath{\left|#1\right>}}

\newcommand{\Rb}{${}^{87}$Rb}

\newcommand{\figref}[1]{Fig.~\ref{#1}}
\newcommand{\Figref}[1]{Figure~\ref{#1}}
\newcommand{\etal}{{\it et al.\/}}
\newcommand{\ie}{{\it i.e.\/}}

\newcommand{\eref}[1]{Eq.~(\ref{#1})}
\newcommand{\Eref}[1]{Equation (\ref{#1})}
\newcommand\diff{\mathop{}\!d}
\newcommand{\ii}{\mathrm{i}}
\newcommand{\ee}{\mathrm{e}}
\newcommand{\dd}{\ensuremath{\mathrm{d}}}

\newcommand{\particleAdd}{}
\newcommand{\bfell}{\mathbf{\bm\ell}}
\renewcommand{\vec}[1]{\ensuremath{\boldsymbol{#1}}}

\begin{document}
\preprint{AIP/123-QED}

\title{Microscopy of an ultranarrow Feshbach resonance using a laser-based atom collider:\\ A quantum defect theory analysis.}

\author{Matthew Chilcott}
\affiliation{
Department of Physics, QSO—Quantum Science Otago, and Dodd-Walls Centre for Photonic and Quantum Technologies,
University of Otago, Dunedin 9016, New Zealand
}%
\author{James F. E. Croft}%
\affiliation{
Department of Physics, QSO—Quantum Science Otago, and Dodd-Walls Centre for Photonic and Quantum Technologies,
University of Otago, Dunedin 9016, New Zealand
}%
\author{Ryan Thomas}%
\affiliation{
Department of Physics, QSO—Quantum Science Otago, and Dodd-Walls Centre for Photonic and Quantum Technologies,
University of Otago, Dunedin 9016, New Zealand
}%
\affiliation{Department of Quantum Science and Technology, Research School of Physics, The Australian National University, Canberra 2601, Australia}

\author{Niels Kj{\ae}rgaard}%
 \email{niels.kjaergaard@otago.ac.nz}
\affiliation{
Department of Physics, QSO—Quantum Science Otago, and Dodd-Walls Centre for Photonic and Quantum Technologies,
University of Otago, Dunedin 9016, New Zealand
}%
\date{\today}

\begin{abstract}
  We employ a quantum defect theory framework to provide a detailed analysis of the interplay between a magnetic Feshbach resonance and a shape resonance in cold collisions of ultracold \Rb{} atoms as captured in recent experiments using a laser-based collider~[\href {https://doi.org/10.1103/PhysRevResearch.3.033209}
  {Phys. Rev. Research {\bf 3}, 033209
  (2021)}]. By exerting control over a parameter space spanned by both collision energy and magnetic field, the width of a Feshbach resonance can be tuned over several orders of magnitude. We apply a quantum defect theory specialized for ultracold atomic collisions to fully describe of the experimental observations. While the width of a Feshbach resonance generally increases with collision energy, its coincidence with a shape resonance leads to a significant additional boost. By conducting experiments at a collision energy matching the shape resonance and using the shape resonance as a magnifying lens we demonstrate a feature broadening to a magnetic width of 8~G  compared to a predicted Feshbach resonance width $\ll 0.1$~mG.
\end{abstract}

\maketitle

\section{Introduction}
Collisional resonances are ubiquitous in atomic and particle physics,
where they arise due to coupling between the scattering continuum and
a quasi-bound state. Their tell-tale signature is an abrupt suppression or
enhancement in scattering as the collision energy is scanned, but they may also emerge when scanning an external field which tunes the energy levels of the system. In ultracold atomic physics, such
field-tunable resonances provide an indispensable tool for
manipulating the interactions between atoms, which has been exploited
in a number of hallmark quantum experiments in atomic systems, including solitons
\cite{Donley2001}, the BEC-BCS crossover~\cite{PhysRevLett.92.040403,
  PhysRevLett.92.120403}, and quantum droplets~\cite{Cabrera301}.

With the recent push towards experiments with ultracold
molecules~\cite{ni.ospelkaus.ea:high,danzl.haller.ea:quantum},
the interaction between collisional resonances has become a subject of
increased interest, due to the high density of states in molecules compared to atoms.
Extraordinarily long lifetimes, approaching milliseconds~\cite{gregory.blackmore.ea:loss,
bause.schindewolf.ea:collisions,gersema.voges.ea:probing},
have been observed in collisions between nonreactive ultracold molecules in
their absolute ground state.
These long lifetime have been attributed to the high density of states of
molecules~\cite{mayle.quemener.ea:scattering,croft.bohn.ea:unified,croft2021anomalous} and
suggest the presence of overlapping resonances~\cite{christianen.groenenboom.ea:lossy}.
Interacting Feshbach resonances are also of importance in collisions of
ultracold magnetic lanthanides, such as erbium and dysprosium, and have been
used to reveal the chaotic nature of the collision process~\cite{frisch.mark.ea:quantum,
durastante.politi.ea:feshbach}.

\begin{figure}
  \centering
  \includegraphics[width=\linewidth]{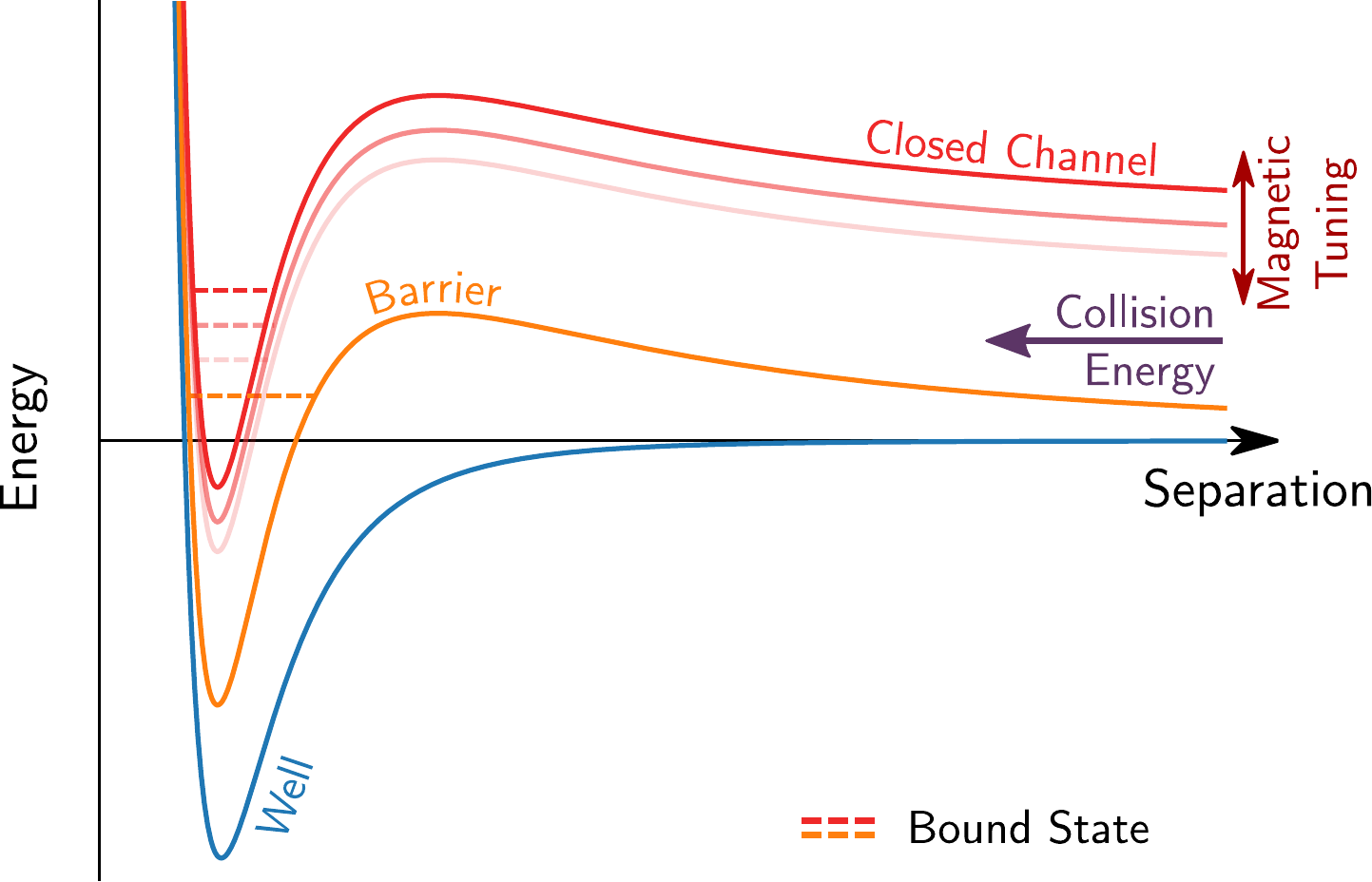}
  \caption{\label{fig:potentials} Representative interaction potentials. The
    $s$-wave channel is shown in blue, and the
    addition of the $d$-wave angular momentum gives rise to the solid
    orange potential, with a barrier hosting a quasi-bound state
    (dashed). The centrifugal barrier is highly exaggerated to be visible---in reality its height is orders magnitudes smaller than the depth of the well. We also show the potential of a closed
    channel potential in red, also with $d$-wave character and hosting a bound state (dashed). The relative
    position of the closed channel can be tuned with a magnetic field,
    and the bound state associated with the Feshbach resonance
    can be pulled through the shape resonance.}
\end{figure}

\red{Multichannel quantum defect theory (MQDT) was originally developed to describe an electron moving in a long-range Coulomb potential (see \cite{seaton:quantum} and the references within). This was later generalized to treat scattering problems involving broader classes of potentials including the long-range van der Walls interaction for atomic collisions~\cite{PhysRevA.19.1485, Mies1980, Fano1982, Mies1984, PhysRevA.58.4222, gao:solutions}. For the low energies characteristic of scattering in the cold and ultracold domain MQDT has proven particularly fruitful, capturing the the physics at threshold~\cite{Gao1996,Burke1998,mies.raoult:analysis,Sadeghpour2000,mies2000,PhysRevA.70.012710,julienne.gao:simple, julienne:ultracold,RevModPhys.82.1225,PhysRevA.88.052701}.}
By separating the scattering problem into energy sensitive and insensitive components, QDT allows for an elegant description of resonance interactions, with the energy dependence encapsulated in a few analytic parameters. Recent demonstrations of the power of QDT include the prediction and interpretation of triplet structures for $d$-wave Feshbach resonances \cite{Cui_2017} and shape resonances \cite{Yao_2019}. 

In this paper, we study the interaction between a shape resonance and a Feshbach resonance in collisions of \Rb{} atoms, with both resonances arising from quasi-bound $d$-wave states of the system.
\Figref{fig:potentials} shows an example potential well for the
collisional entrance channel (blue line) which describes the interaction as a function of radial separation. When including angular momentum, the effective potential contains a barrier in front of the potential well (orange line), behind which a quasi-bound state can be formed (orange dashed line). For atoms tunnelling through the barrier, such a quasi-bound state gives rise to a scattering resonance. Similarly, Feshbach resonances arise from the coupling to a bound state, but in this case the bound state belongs to a closed channel. A closed channel potential, which is energetically inaccessible for separated particles, is presented in red in
\figref{fig:potentials}. Due to the deep potential well, the
closed channel shown becomes accessible at
\blaa{short range} during a collision. If a bound state (red dashed line) is present at the collision energy, incoming atoms can temporarily bind in this state, enhancing their interaction. In the
case of a magnetic Feshbach resonance, the
different magnetic moments of the closed and entrance channels allows us to
manipulate the position of the Feshbach resonance in energy.

While shape resonances and Feshbach resonances are typically treated in isolation, the tunability of the latter opens up the possibility of moving a Feshbach resonance through a shape resonance using an external field. 
We recently studied the interaction of a pair of such resonances and observed the avoided crossing in the associated $\bm S$ matrix poles~\cite{Chilcott2021A}. In
the current work, we \red{revisit the data acquired in these experiments and} extend our analysis of this resonance pair \red{to} give an interpretation in terms of multi-channel quantum defect theory (MQDT). 
We show that a simple two-channel QDT model captures all the essential physics of the interacting collisional resonances.

\section{Experimental Methods}\label{sec:experimental}
\subsection{System}
We collide pairs of \Rb{} atms which are both in the absolute ground hyperfine state $\ket{F, m_F} = \ket{1,1}$. This $\ket{1,1}\particleAdd\ket{1,1}$
entrance channel has a plethora of Feshbach resonances, previously
mapped out by loss spectroscopy~\cite{PhysRevLett.89.283202}. In the
current study, we utilize a $d$-wave Feshbach resonance corresponding to the closed channel
molecular state with the quantum numbers $F_1 = 2, F_2 =
2, v' = -5$, and $M = 2$, where $M = m_{F_1} + m_{F_2}$ and $v'$ is the
vibrational quantum number counting from the $F_1 = 2, F_2 = 2$
threshold. For a magnetic field of \SI{930}{\gauss}, this state is located at the $\ket{1,1}\particleAdd\ket{1,1}$-channel threshold where it is predominantly comprised of
components from the $\ket{2,0}\particleAdd\ket{2,2}$ and $\ket{2,1}\particleAdd\ket{2,1}$
channels. The $\ket{1,1}\particleAdd\ket{1,1}$ channel
also hosts a $d$-wave shape resonance at a collision energy  near \SI{300}{\micro\kelvin}
\cite{PhysRevLett.93.173201,PhysRevLett.93.173202}, as measured in units of the Boltzmann constant $k_{\rm B}$.

The Feshbach resonance we employ was predicted by Marte \etal~\cite{PhysRevLett.89.283202} to be
located at \SI{930.9}{\gauss} with a theoretical width of
$\ll\SI{0.1}{\milli\gauss}$; their experiments observed it at
$B_0=\SI{930.02}{\gauss}$ using loss spectroscopy. A subsequent observation has placed this
resonance at \SI{930.89}{\gauss} in photo-association experiments
\cite{Eisele2021}. Our own loss spectroscopy measurements (described
in Appendix~\ref{app:loss}) observe the zero-energy resonance at
\SI{929.921(3)}{\gauss}.

\subsection{Optical collider}
Our collider is composed of a system of steerable optical dipole
traps, formed by pairs of crossed, red-detuned laser beams
\cite{chisolm2018}. The procedure to prepare two ultracold ($\sim\SI{800}{\nano\kelvin}$) $\ket{1,1}$-state \Rb{} clouds in separate crossed dipole traps is detailed in \cite{Chilcott2021A}.

We tune the position of the Feshbach resonance by applying a magnetic field with a
pair of Helmholtz coils carrying a current controlled at the ppm level
\cite{Thomas2020}, before accelerating clouds each containing $\sim3\times10^5$ into collision at specific energies in the range \SIrange{156}{850}{\micro\kelvin}. The acceleration is achieved by steering the laser trapping beams and as the clouds reach the collision
energy, all laser beams are turned off so that the atoms collide in the
absence of trapping.

\begin{figure}
  \centering
  \includegraphics[width=\linewidth]{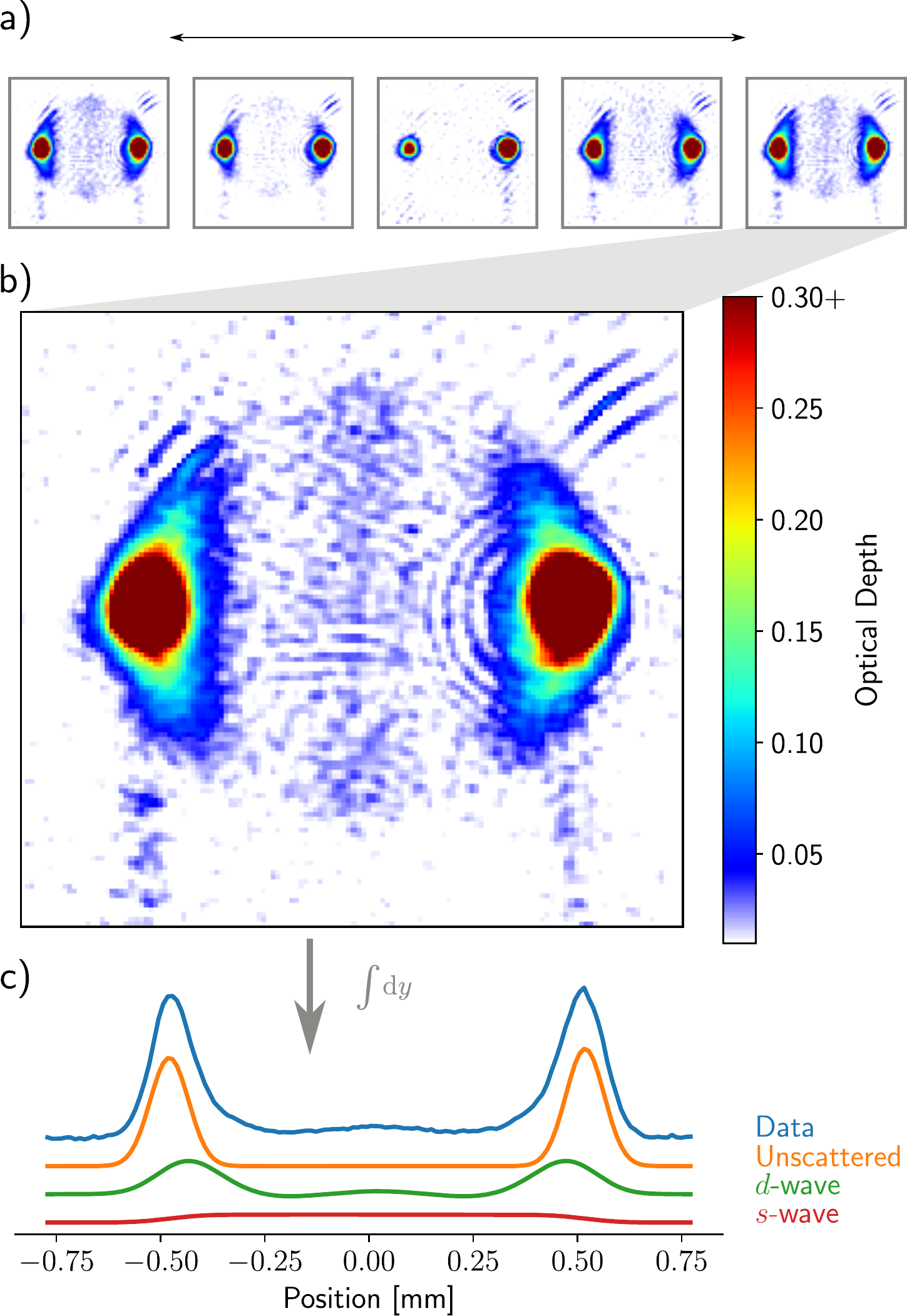}
  \caption{\label{fig:absimg} Fractions of scattered atoms from post-collision absorption images. a) Absorption mages at a range of magnetic fields over the resonance for a collision energy of $E/k_{\rm B}=\SI{270}{\micro\kelvin}$. The collision axis is horizontal. b) Close-up of a collision near \SI{940}{\gauss}. The
    scattering here is strongly anisotropic and has the signature of $d$-wave scattering \cite{Kjaergaard_2004}.
    c) The integrated column density (blue), to which our model is fitted
    in order to extract the parameters of the collision. From the fit the components of unscattered (orange), and the $s$-wave (red) and $d$-wave (green) scattered atoms are extracted.}
\end{figure}
\subsection{Detection}\label{sec:det}
After the clouds separate post-collision, we acquire an absorption image of the clouds and the halo of scattered atoms. \Figref{fig:absimg}a
shows examples of such images  with the atom distribution projected
onto a plane and exhibiting clear $d$-wave character
\cite{PhysRevLett.93.173201,Kjaergaard_2004}. We integrate the an image (\figref{fig:absimg}b) in the
direction orthogonal to the collision (vertically) and the
resulting integral (\figref{fig:absimg}c) carries a spatial imprint of shapes associated
with the interfering $s$ and $d$ partial waves, and the unscattered thermal
clouds. By fitting these shapes to the integrated image, we extract
the scattered fraction $\mathcal{S}$.

The cross-section is related to the scattered fraction by~\cite{Thomas2018}
\begin{equation}
  \mathcal{S} = \frac{\alpha\sigma}{1 + \alpha\sigma},
\end{equation}
where $\sigma = \sigma_{s} + \sigma_{d}$ is the sum of the $s$ and $d$
partial-wave cross-sections, and the parameter $\alpha$ is geometry-dependent
and left as a free parameter when fitting the
cross-section. For a particular partial wave $\ell$, the cross-section is
\begin{equation}\label{eqn:crosssection}
  \sigma_\ell = \frac{4\pi\hbar^2(2\ell +
    1)}{\mu E}\sin^2\big(\delta_\ell\big),
\end{equation}
where $\delta_{\ell}$ is the corresponding partial wave scattering phase shift, $E$
is the collision energy, and $\mu$ is the reduced mass. As we
are colliding indistinguishable bosons, only even-$\ell$ partial waves are allowed and \eref{eqn:crosssection} includes an
additional factor of 2 compared to the distinguishable particle case. Because the magnetic Feshbach resonance has a 
$d$-wave character, we take the $s$-wave phase-shift $\delta_s$, and
consequently $\sigma_s$ to be constant in magnetic field and only a
function of energy.  Close to the magnetic resonance, the $d$-wave
phase-shift $\delta_{d}$ can be described by the Beutler-Fano model:
\begin{equation}\label{eqn:FeshbachPhase}
\delta_d(E,B) = \delta_{\text{bg}}(E) +
  \arctan\left(\frac{\Gamma_B(E) / 2}{B -
    B_\text{res}(E)}\right),
\end{equation}
for a magnetic field $B$, and a collision energy $E$.

With this resonance model, we extract the $d$-wave background phase shift
$\delta_\text{bg}(E)$, along with the width $\Gamma_B(E)$, and position $B_\text{res}(E)$
of the Fano lineshape by sweeping the magnetic field at constant energy.

\begin{figure}
  \centering
  \includegraphics[width=\linewidth]{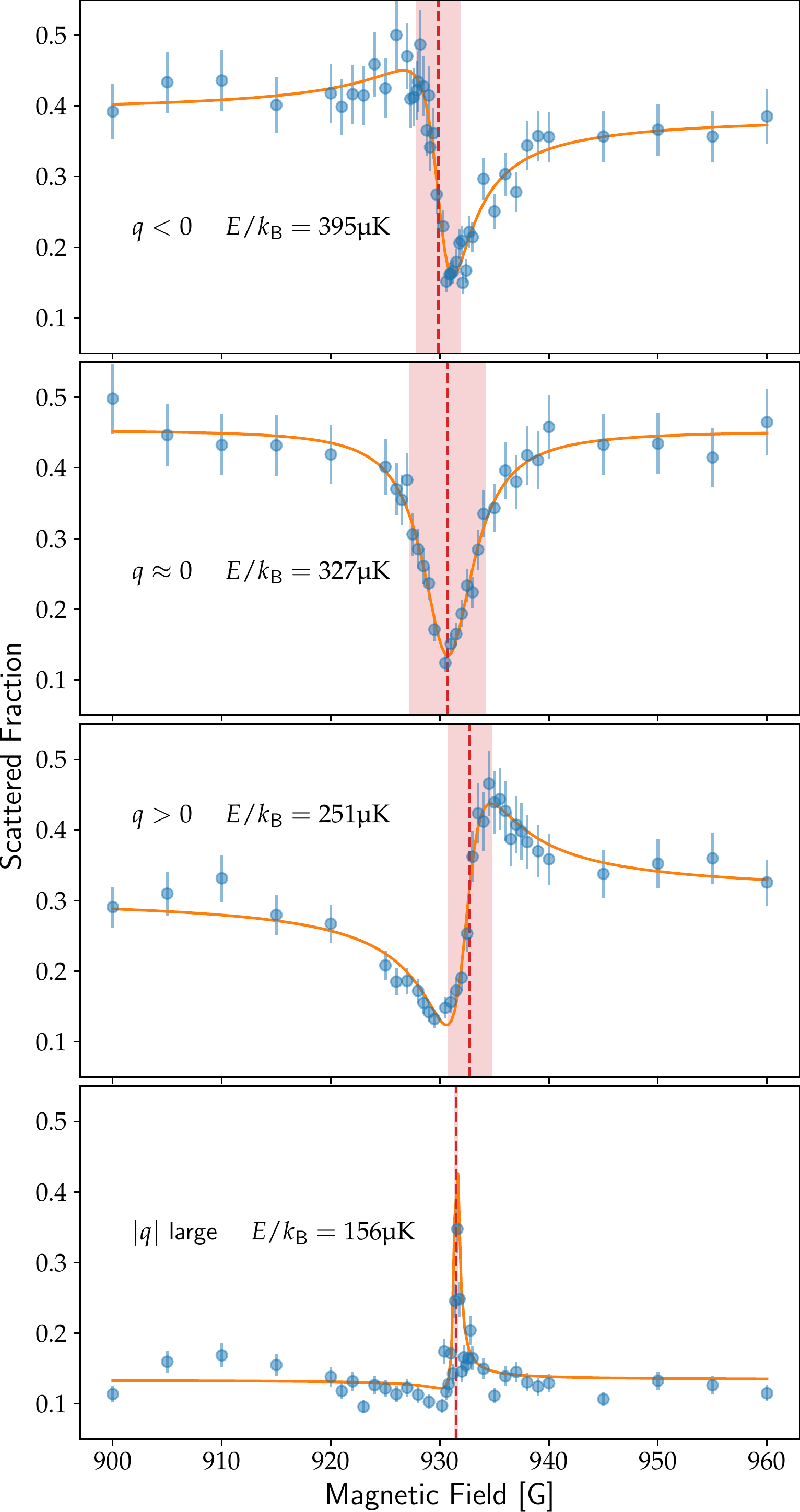}
  \caption{\label{fig:fano} Examples of Fano profiles of differing shape parameters $q$ as encountered in our experiments. Points shows the measured scattered
    fraction as a function of magnetic field for four different collision energies $E$.  The solid lines are curve fits based on the Beutler-Fano model outlined in section \ref{sec:det}.
    The fitted parameters $B_{\rm res}$ and $\Gamma_B$ [cf. \eref{eqn:FeshbachPhase}] are indicated as a dashed line and and a shaded area, respectively. The experimental data illustrating the characteristic Fano profiles  have previously been published in \cite{Chilcott2021A}}
\end{figure}

The Beutler-Fano model above is equivalent to the common form of the
Fano profile~\cite{RevModPhys.82.2257},
\begin{equation}\label{eqn:fano}
  \sigma \propto \frac{(q + \epsilon)^2}{1 + \epsilon^2},
\end{equation}
where $q$ is the so-called shape parameter and $\epsilon$ is the
scaled, dimensionless parameter in which the resonance occurs.
In our case, the required mapping to this form is given by $q =
\cot(\delta_\text{bg})$ and $\epsilon = 2(B - B_\text{res})/\Gamma_B$,
the inverse of the $\arctan$ argument in \eref{eqn:FeshbachPhase}. 
Representative Fano profiles measured at four different energies are shown in
\figref{fig:fano}, illustrating different regimes of the $q$ parameter.
The line shape parameterized by $q$ can be thought of as due to the interference between the two pathways, 
the asymmetric line shape is then arises due to the constructive interference on one side and destructive on the other~\cite{fano.rau:atomic,Rau_2004}.
 
We note that the limiting cases provide a symmetric dip at $q = 0$, and a
Lorentzian profile at $|q| \to \infty$, while intermediate values of $q$ are tied to asymmetric profiles with a `polarity' determined by the sign of $q$.
\section{Quantum Defect Theory}
The compelling variation in the Fano profiles observed in \figref{fig:fano} results from the interplay between the Feshbach resonance and a shape resonance. We employ quantum defect theory (QDT) to characterize and interpret the intriguing resonant scattering behavior.
\red{As mentioned in the introduction QDT is a well-developed theory. 
There are however different notations (most notably those of Greene, Rau, and Fano~\cite{Fano1982} and Mies~\cite{Mies1984}) and approximations employed by various QDT treatments.
As such, here we provide a self contained treatment which combines the numerically stable approach of Ruzic 
\textit{et al.}~\cite{ruzic.greene.ea:quantum} with a reference function optimisation~\cite{giusti-suzor.fano:alternative}.
The importance of optimized reference functions in the analysis of Feshbach resonances in ultracold atomic collisions has previously been discussed by Oss{\'{e}}ni \textit{et al}.~\cite{osseni.dulieu.ea:optimization}.}
\begin{figure*}
\includegraphics[width=\linewidth]{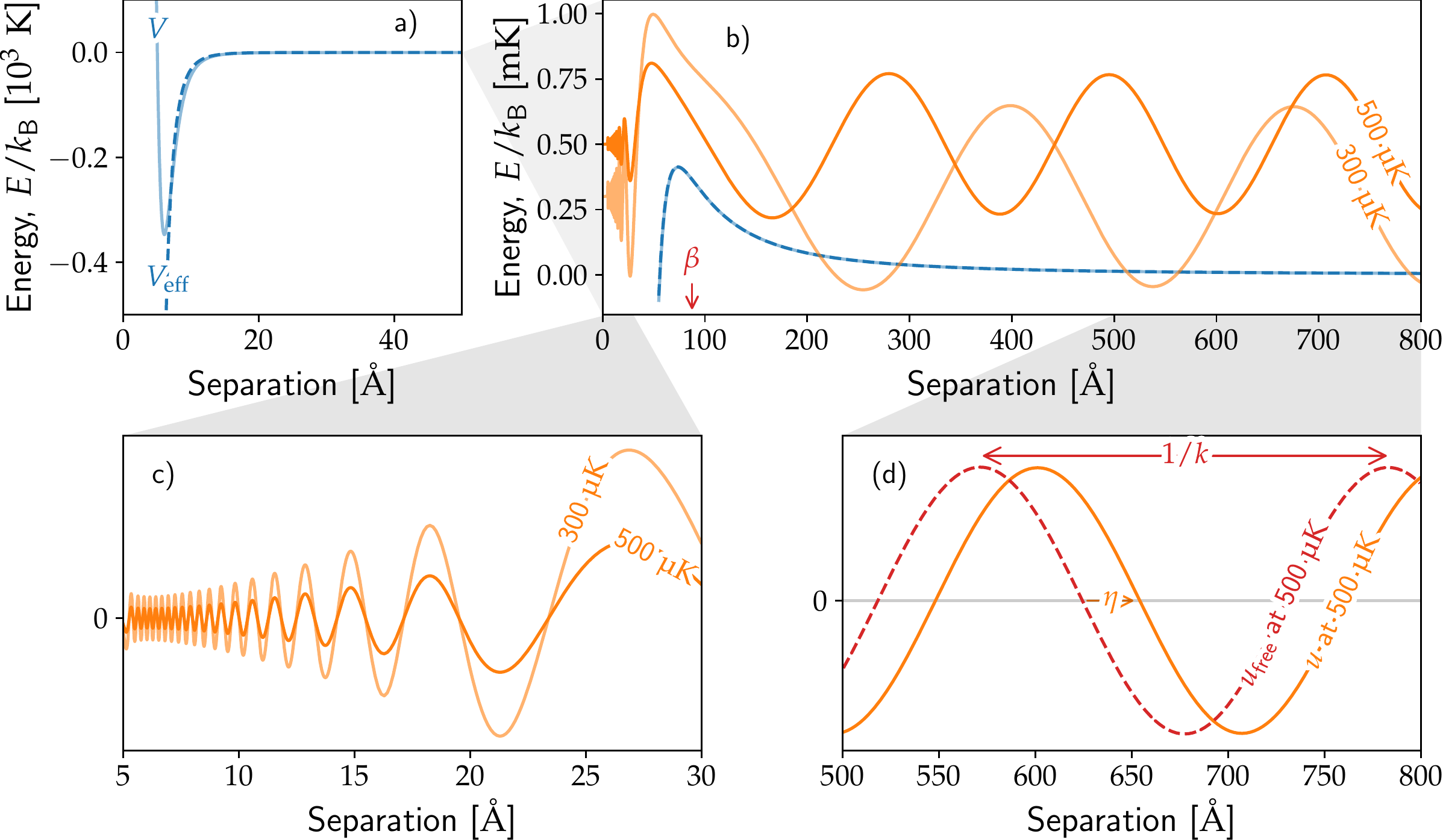}
\caption{The different length and energy scales of the scattering problem in the $\ell = 2$ entrance channel. (a) The triplet potential of the entrance channel, $V$, from Ref.~\cite{Pashov2007}. At the long-range, $V$ is well-approximated by the van der Waals potential $V_\text{eff}$. (b) At threshold and over a wider range of separation, we show two scattering wavefunctions derived from coupled-channels calculations, offset vertically by their collision energies.  An arrow labels the length scale of the van der Waals potential, $\beta$. (c) At \blaa{short range}, the shape of the wavefunction is independent of energy apart from variations of the amplitude. The increase in amplitude demonstrated at \SI{300}{\micro\kelvin} is associated with the nearby shape resonance. (d) A scattering wavefunction $u$ at long range along with the free-particle wavefunction $u_\text{free}$. The difference between the two is captured by a scattering phase-shift $\eta$.\label{fig:waves}}
\end{figure*}
\subsection{Overarching QDT framework}
The QDT approach to cold collisions takes advantage of the
separation of energy and length scales common in scattering problems~\cite{seaton:quantum,Fano1982,Mies1984,
mies.raoult:analysis,PhysRevA.70.012710}. 
The solid blue line in \figref{fig:waves}a,b shows the triplet interaction potential between two \Rb{} atoms as a function of separation. Asymptotically, this potential goes to zero. Atoms at \blaa{short range} (\figref{fig:waves}a) encounter a potential with a depth on the order of a thousand Kelvin, compared to a collision energy of the order of a millikelvin. This separation of energy and length scales allows the important physics of collisions to be captured by a few analytic
parameters~\cite{mies.julienne:multichannel,julienne.gao:simple, julienne:ultracold,RevModPhys.82.1225,PhysRevA.88.052701}. Importantly, computing these QDT parameters for a system only requires the long-range form of the interaction potential and the masses of the atoms.

To set the scene, we show in \figref{fig:waves}b the energy-normalised scattering wave functions in the $\ket{1,1}\particleAdd\ket{1,1}$ channel, generated by full coupled-channel calculations \footnote{The coupled-channels calculations, as used in Ref.~\cite{Chilcott2021A}, are generated by propagating the log-derivative of the wavefunction using the technique of Ref.~\cite{Manolopoulos1986}, and extracting the wavefunction using the technique of Ref.~\cite{Thornley1994}} \nocite{Manolopoulos1986,Thornley1994}at two different energies ($E/k_\text{B}=\SI{300}{\micro\kelvin}$ and $E/k_\text{B}=\SI{500}{\micro\kelvin}$). These wave functions highlight two salient features shared by all scattering solutions. Firstly, inside the well solutions are oscillatory with a varying local wave number which, crucially, has virtually no dependence on the collision energy $E$; the common waveforms differ only in amplitude (see \figref{fig:waves}c). This difference in amplitude arises due to the asymptotic energy normalization. 
Secondly, the long-range asymptotic solutions are sinusoidal with a constant wave number. This wave number will depend strongly, but trivially, on the collision energy $E$ as $k=\sqrt{2\mu E/\hbar^2}$ and for a given energy the sinusoid will display a phase shift $\eta(E)$ with respect to the regular free space solution at that energy (\figref{fig:waves}d). Because the specific wave functions in  \figref{fig:waves} satisfy the physical boundary conditions of the scattering problem, the partial (elastic) scattering cross section is determined by this phase shift using \eref{eqn:crosssection}, with $\delta_\ell = \eta$.
By observing that all scattering solutions share some common ground via the above two highlighted features, the QDT framework will enable us to predict the energy dependence of the phase shift analytically.
\subsubsection{Coupled channels equations, K matrix and S matrix}
The objective of any scattering problem is to obtain the system's $\bm S$ matrix, as it completely describes the outcome of a scattering experiment. Formally, the $\bm S$ matrix can be constructed from computed wavefunctions, which are solutions to the coupled radial Schr\"{o}dinger equations,
\begin{widetext}
\begin{equation}\label{eqn:qdt:coupl}
	\sum_{j=1}^{N}\left\{\left[ -\frac{\hbar^2}{2\mu}\frac{\dd^2}{\dd r^2}+\frac{\ell(\ell+1)\hbar^2}{2\mu r^2}\right]\delta_{ij}\\+V_{ij}(r)\right\}u_j(r)=E_iu_i(r),
\end{equation}
\end{widetext}
where the energy $E_i$ for a channel $i$ is measured with respect to its threshold and $V_{ij}$ are the elements of a potential matrix. In our experiments, we consider two atoms entering as $\ket{1,1}\particleAdd\ket{1,1}$. If we ignore the weak spin-spin dipole interactions, collisions between the atoms $1$ and $2$ conserve
$m_{F_1}+m_{F_2}=2$ as well as mechanical angular momentum $\ell$~\cite{Tiesinga1996}. For the $d$-wave channel, their coupling is restricted to the 
channels $\ket{1,1}\particleAdd\ket{2,1}$, $\ket{1,0}\particleAdd\ket{2,2}$,
$\ket{2,0}\particleAdd\ket{2,2}$, and $\ket{2,1}\particleAdd\ket{2,1}$ with $\ell=2$, which results in an $N = 5$-channel set of equations~\cite{mies2000}. 
Designating the entrance channel $\ket{1,1}\particleAdd\ket{1,1}$ with $i=1$, we have $E_1>0$ as the asymptotically free incoming $\ket{1,1}$ atoms have a non-zero relative kinetic energy. In our experiment the collision energy is so low that $E_i<0$ for the remaining four channels; these are all energetically closed and atoms can only leave the collision via the $i=1$ channel.

Generally, when solving coupled channels problems like \eref{eqn:qdt:coupl}, one typically propagates out an $N\times N$ regular solution matrix $\bm F$~\footnote{In practice, for reasons of numerical stability, it is more common to propagate the log-derivative of this matrix} where $N=N_\text{o}+N_\text{c}$ is the sum of the number of open and closed channels. The $N$ columns of $\bm F$ represent $N$ linearly independent solution vectors to \eref{eqn:qdt:coupl} with row $i$ of a solution vector corresponding to the channel $i$. Asymptotically, the solution matrix can be decomposed as \cite{Hutson2009} 
\begin{eqnarray}\label{eqn:cc1}
{\bm F}(r)&  \overset{r \to \infty}\sim&
\boldsymbol{\mathcal{J}}(r)+\boldsymbol{\mathcal{N}}(r){\bm K},
\end{eqnarray}
where ${\bm K}$ is a constant real symmetric matrix, and $\boldsymbol{\mathcal{J}}$ and $\boldsymbol{\mathcal{N}}$ are diagonal matrices with entries
\begin{subequations}
\begin{eqnarray}
	\mathcal{J}_{ii}(r)
	&=& \begin{cases}
	 rk_i^{1/2}j_{\ell}(k_ir),\,\text{for $i$ open} \\
	(k_ir)^{-1/2}\mathcal{I}_{\ell+1/2}(k_ir),\,\text{for $i$ closed}
	\end{cases},
\end{eqnarray}
\begin{eqnarray}
	\mathcal{N}_{ii}(r)
	&=& \begin{cases}
	 rk_i^{1/2}n_{\ell}(k_ir),\,\text{for $i$ open} \\
	(k_ir)^{-1/2}\mathcal{K}_{\ell+1/2}(k_ir),\,\text{for $i$ closed}
	\end{cases}.
\end{eqnarray}
\end{subequations}
Here, $j_\ell$ and $n_\ell$ are spherical Bessel functions of the first and second kind, respectively, and $\mathcal{I}_{\ell+1/2}$ and $\mathcal{K}_{\ell+1/2}$ are modified Bessel functions of the first and third kind, respectively.

The energy dependent $\bm K$ matrix defined by \eref{eqn:cc1} contains all the scattering behavior of the system. In particular, the $N_\text{o}\times N_\text{o}$ sub-matrix of $\bm K$ that pertains to only the open channels defines the $\bm S$ matrix \cite{Hutson2009,Burke2013,Friedrich2015}:
\begin{equation}\label{eqn:K_to_S_normal}
  \bm S = \left(1 + \ii{\bm K}_\text{oo}\right){\left(1 - \ii{\bm K}_\text{oo}\right)}^{-1}.
\end{equation}
We note that ${\bm K}_\text{oo}$ is known as the reactance matrix in the literature.
In the treatment of identical particles, the wave functions must be properly symmetrized~\cite{PhysRevB.38.4688}. Because we only consider elastic collisions, we simply need to include a factor of 2 in the even-$\ell$ partial-wave cross sections, \textit{cf.}~\eref{eqn:crosssection}.

\subsubsection{QDT treatment: uncoupled channels at long range}
Rather than directly numerically solving \eref{eqn:qdt:coupl} to, in turn, obtain $\bm K$ and $\bm S$, QDT proceeds by assuming that beyond a certain separation, $R_{\rm int}$, all channels are uncoupled.
\red{Outside this distance, the radial wave function for the entrance channel is therefore a solution to the radial Schr\"odinger equation},
\begin{equation}\label{eqn:qdt:long_range}
	\left[ -\frac{\hbar^2}{2\mu}\frac{\dd^2}{\dd r^2}+\frac{\ell(\ell+1)\hbar^2}{2\mu r^2}+V_1(r)\right]u_1(r)=E_1 u_1(r).
\end{equation}
For our system, the long-range behaviour of $V_1$ is well-described by a Van der Waals potential, $V_1(r)=-C_6/r^6$, and \eref{eqn:qdt:long_range} takes the specific form
\begin{equation}\label{eqn:qdt:long_range2}
\frac{\dd^2}{\dd R^2}u_1(R)=\left(\underbrace{-\frac{1}{R^6}+\frac{\ell(\ell+1)}{R^2}}_{V_{\rm eff}(R)}-\bar{E}_1\right)u_1(R),
\end{equation}
where the radial distance $R$ is measured in units of the van der Waals radius, $\beta\equiv (2\mu C_6/\hbar^2)^{1/4}$ and energy on a scale $E_\beta\equiv\hbar^2/2\mu\beta^2$.
We also define the local wave number
\begin{equation}\label{wavenumber}
\bar{k}_1(R)=\sqrt{\bar{E}_1-V_{\rm eff}(R)}.
\end{equation}

For reference, we note that our particular system has $C_6=\SI{3.253e7}{\kelvin\angstrom\tothe{6}}$
for \Rb{}~\cite{derevianko.babb.ea:high-precision}, which gives $\beta=\SI{87.37}{\angstrom}$ and $E_\beta = \SI{73.11}{\micro\kelvin}$.
\subsection{QDT reference functions}
Knowing that our long-range behaviour is well described by the van der Waals potential, we compute the QDT reference functions in this potential.

\subsubsection{Asymptotic considerations (\texorpdfstring{$R \to \infty$}{R to infinity})} \label{sec:long}
\red{As $R \to \infty$}, $V_\text{eff}$ vanishes and the $\bar{E}_1$-term on the right hand side of \eref{eqn:qdt:long_range2} dominates. In this region an energy-normalized solution $u_1(R)$ is sinusoidal, oscillating with the asymptotic wavenumber $\bar{k}_1 \overset{R \to \infty}{=} \sqrt{\bar{E}_1}$:
\begin{eqnarray}\label{eqn:qdt:long_range3}
u_1(R)&  \overset{R \to \infty}\sim& \frac{1}{\sqrt{\bar{k}_1}}\sin(\bar{k}_1R-\ell\pi/2+\eta_1),
\end{eqnarray}
where the $\ell\pi/2$-term references the phase shift $\eta_1$ against the regular free particle solution for the given partial wave,
\begin{eqnarray}\label{eqn:qdt:long_range4}
u_{\rm free}(R)=\bar{k}_1R j_{\ell}(\bar{k}_1R)&  \overset{R \to \infty}\sim&\sin(\bar{k}_1R-\ell\pi/2).
\end{eqnarray}
Equivalently, \eref{eqn:qdt:long_range3} can be written as
\begin{eqnarray}\label{eqn:qdt:long_range5}
u_1(R)&  \overset{R \to \infty}\sim&
c_1\frac{1}{\sqrt{\bar{k}_1}}\sin\left(\bar{k}_1R-\ell\pi/2+\xi_1\right) \nonumber\\ & & +c_2\frac{1}{\sqrt{\bar{k}_1}}\cos\left(\bar{k}_1R-\ell\pi/2+\xi_1\right),
\end{eqnarray}
where the coefficients for the two quadrature components are given by 
\begin{equation}
    \begin{bmatrix}
c_1\\
c_2
\end{bmatrix}= \begin{bmatrix}
\sin\xi_1 & \cos\xi_1\\
(-1)^{\ell}\cos\xi_1 & (-1)^{\ell+1}\sin\xi_1
\end{bmatrix} \begin{bmatrix}
\sin\eta_1\\
\cos\eta_1
\end{bmatrix},
\end{equation}
for a particular choice of the arbitrary phase $\xi_1$. 
Guided by \eref{eqn:qdt:long_range5}, the solution of \eref{eqn:qdt:long_range2} can be expressed as
\begin{equation}\label{sollong}
u_1(R)= c_1f_1(R)+c_2g_1(R), 
\end{equation}
where $f_1$ and $g_1$ are exact solutions to \eref{eqn:qdt:long_range2} defined by the boundary conditions 
\begin{subequations}\label{eqn:qdt:longrange}
\begin{eqnarray}\label{eqn:qdt:long_range6}
f_1(R)&  \overset{R \to \infty}\sim&
\frac{1}{\sqrt{\bar{k}_1}}\sin\left(\bar{k}_1R-\ell\pi/2+\xi_1\right),
\end{eqnarray}
and
\begin{eqnarray}\label{eqn:qdt:long_range7}
g_1(R)&  \overset{R \to \infty}\sim&
\frac{1}{\sqrt{\bar{k}_1}}\cos\left(\bar{k}_1R-\ell\pi/2+\xi_1\right).
\end{eqnarray}
\end{subequations}
In \eref{sollong} the coefficients $c_1$ and $c_2$ depend on the choice of the free parameter $\xi_1$. In particular, for $\xi_1=\eta_1$ $c_1=1$ and $c_2=0$.
\subsubsection{Considerations inside the van der Waals radius (\texorpdfstring{$R \lesssim 1$}{R << 1})}\label{sec:short}
\begin{figure}
  \includegraphics[width=\linewidth]{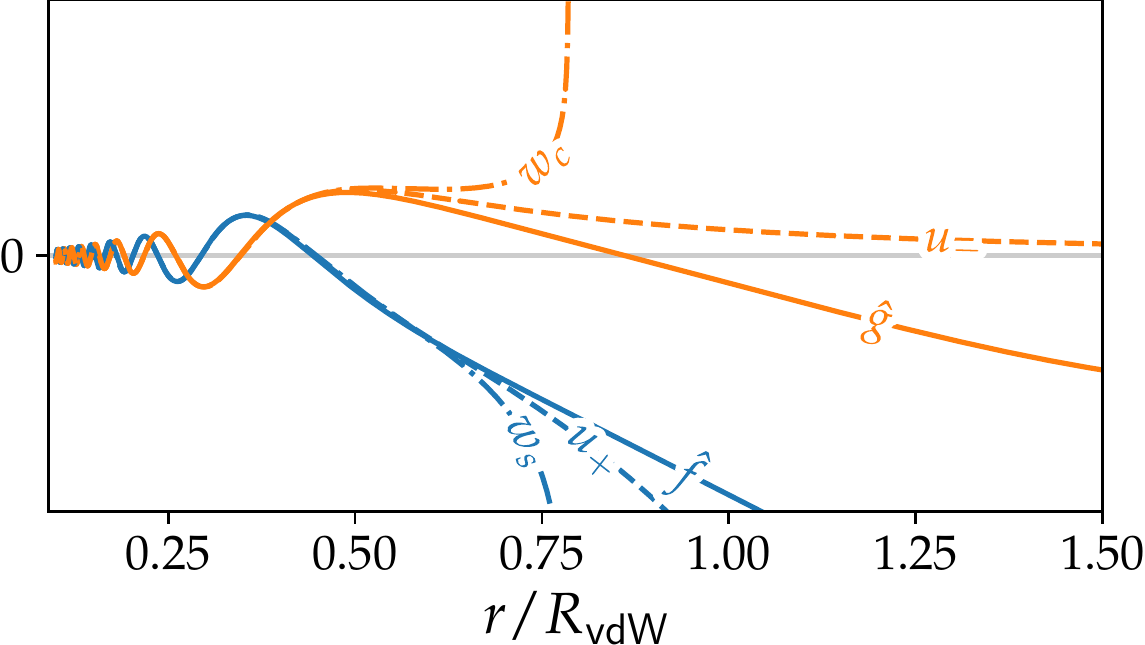}
  \caption{Solutions  pairs inside the van der Waals radius: the zero-energy analytic solutions $u_\pm$, the WKB solutions $w_s,\,w_c$ for $\phi_1 = 0$ (computed using $R_0=0.1$) at $E/k_{\rm B}=$\SI{300}{\micro\kelvin}, and the corresponding reference functions $\hat{f}_1,\,\hat{g}_1$. 
  \label{fig:wavefunction}}
\end{figure}

For $R \lesssim 1$, the $\bar{E}_1$-term of \eref{eqn:qdt:long_range2} is negligible and the solution $u_1(r)$ becomes reminiscent of the solution $u^{(0)}_1(r)$ of
the Bessel equation
\begin{equation}\label{eqn:qdt:short_range8}
\frac{\dd^2}{\dd R^2}u^{(0)}_1(R)=-\left[\frac{1}{R^6}-\frac{\ell(\ell+1)}{R^2}\right]u^{(0)}_1(R).
\end{equation}
$u^{(0)}_1(r)$ can be expressed as a linear combination of
\begin{subequations}
\begin{eqnarray}
	u_+(R)&\equiv&\frac{\sqrt{\pi}}{2}\sqrt{R}Y_{(2\ell+1)/4}\left(\frac{1}{2R^2}\right),
\end{eqnarray}
and
\begin{eqnarray}
	u_-(R)&\equiv&\frac{\sqrt{\pi}}{2}\sqrt{R}J_{(2\ell+1)/4}\left(\frac{1}{2R^2}\right),
\end{eqnarray}
\end{subequations}
where $J$ and $Y$ are Bessel functions of the first and second kind, respectively. 
In \figref{fig:wavefunction} we plot $u_\pm(R)$ for $\ell=2$ and it can be seen how $u_-$ decays as $R$ increases, while $u_+$ blows up. This functional behavior is also captured by the limiting forms \cite{Abramowitz1970}
\begin{subequations}
\begin{eqnarray}
	u_+(R)
	&\sim& \begin{cases}
 R^{3/2}\sin\left(\frac{1}{2R^2}-\tfrac{2\ell+3}{8}\pi\right),\text{for }R \rightarrow 0\\
	\frac{\sqrt{\pi}}{2}	R^{\ell+1},\text{for } R\rightarrow\infty
	\end{cases},\phantom{hhsdf}
\end{eqnarray}
\begin{eqnarray}
	u_-(R)
	&\sim& \begin{cases}
	 R^{3/2}\cos\left(\frac{1}{2R^2}-\tfrac{2\ell+3}{8}\pi\right),\text{for }R \rightarrow 0\\
	\frac{\sqrt{\pi}}{2}	R^{-\ell},\text{for } R\rightarrow\infty
	\end{cases}.\phantom{hhsdf}
\end{eqnarray}
\end{subequations}
Both $u_-(R)$ and $u_+(R)$ are exact solutions to \eref{eqn:qdt:short_range8}, \ie, \eref{eqn:qdt:long_range2} with $\bar{E}_1=0$, valid for all $R$. For $E>0$, a pair of linearly independent approximate WKB solutions to \eref{eqn:qdt:long_range2} around some point $R_0$ where the potential $V_{\rm eff}$ is deep are given by
\begin{subequations}\label{eqn:qdt:shortrange9i}
\begin{align}\label{eqn:qdt:shortrange9}
  w_s(R)&= \frac{1}{\sqrt{\bar{k}_1(R)}} \sin\left[\theta_\text{WKB}(R)\right], \\
 w_c(R)  &= \frac{1}{\sqrt{\bar{k}_1(R)}} \cos\left[\theta_\text{WKB}(R) \right]\label{eqn:qdt:shortrange9b},
\end{align}
\end{subequations}
where
\begin{equation}\label{eqn:qdt:WKBPhase}
    \theta_\text{WKB}(R) = -\int^R_{R_0} \diff R' \bar{k}_1(R') + \frac{1}{2}R_0^{-2} - \frac{2\ell+3}{8}\pi+ \phi_1.
\end{equation}
We note that in the vicinity of $R_0$ (\ie, for values of $R$ where the potential well is deep), $w_s(R)$ and  $w_c(R)$ are largely insensitive to the channel energy $\bar{E}_1$ and that for $\phi_1=0$, they match the analytic zero-energy solutions $u_{\pm}(R)$ in this region (\figref{fig:wavefunction} shows $w_s$ and $w_c$ for $E/k_{\rm B}=300~\mu$K).
The phase $\phi_1$ is the key that unlocks QDT's use of only the long-range potential because it can encapsulate the effects of the complicated multi-channel short-range interaction as a scalar quantity which varies only weakly with energy and hence can be taken to be constant.

Analogous to \eref{sollong}, the solution $u_1(R)$ to  \eref{eqn:qdt:long_range2} can be expressed as the linear combination
\begin{equation}\label{blah}
u_1(R)= \hat{c}_1\hat{f}_1(R)+\hat{c}_2\hat{g}_1(R), 
\end{equation}
where
 $\hat{f}_1$ and $\hat{g}_1$ are exact solutions to \eref{eqn:qdt:long_range2} defined by the initial values
\begin{subequations}\label{eq:two}
\begin{eqnarray}\label{eqn:qdt:long_range8}
\{\hat{f}_1(R_m)=w_s(R_m),
\hat{f}_1'(R_m)=w_s'(R_m)\},
\end{eqnarray}
\begin{eqnarray}\label{eqn:qdt:long_range9}
\{\hat{g}_1(R_m)=w_c(R_m),
\hat{g}_1'(R_m)=w_c'(R_m)\},
\end{eqnarray}
\end{subequations}
where $R_m>R_{\rm int}$ is some point within the potential well (possibly, but not necessarily $R_0$), and the coefficients $\hat{c}_1$ and $\hat{c}_2$ are determined by the boundary conditions of the physical problem. Individually, $\hat{f}_1$ and $\hat{g}_1$ inherit the energy-insensitivity of $w_s$ and $w_c$ inside the potential well.

Figure~\ref{fig:wavefunction} shows that $\hat f$ and $\hat g$ are perfectly tailored to represent the short-range multichannel wave function.
Over the range of collision energies of interest they are essentially independent of energy, and the similarity in the wave functions at short range extends analytically to energies below threshold.
The functions $w_s$ and $\hat{f}_1$, which are plotted in \figref{fig:wavefunction} for an energy $E/k_{\rm B}=$\SI{300}{\micro\kelvin}, are both completely equivalent to the zero-energy solution $u_+$ at short-range; likewise, $w_c$, and $\hat{g}$ are completely equivalent to the zero-energy solution $u_-$.
We also note that, by construction, for $\phi_1=0$, $\hat{g}_1$ links up to the purely decaying zero-energy solution to~\eref{eqn:qdt:short_range8}, namely $u_-$.
This choice provides numerically stable reference functions as it corresponds to propagating the maximally linearly independent pair~\cite{ruzic.greene.ea:quantum}.
\subsubsection{Relating reference functions}
The considerations of sections \ref{sec:long} and \ref{sec:short} provide the two pairs of reference functions $\{f_1(R),g_1(R)\}$ and $\{\hat{f}_1(R),\hat{g}_1(R)\}$, respectively. These are all defined for all $R > R_m$, but while $f_1,$ and $g_1$ refer to asymptotic boundary conditions at long-range, $\hat{f}_1$ and $\hat{g}_1$ refer to boundary conditions within the van der Waals radius. Connecting the pairs of reference functions defines the QDT parameters which underpin the QDT framework.

Like $f_1(R)$, the solution $\hat{f}_1(R)$ [subject to \eref{eqn:qdt:long_range8} and tied to the phase $\phi_1$] will be sinusoidal for $R\rightarrow \infty$.
The phases of the two wave forms can be made to match through the free phase $\xi_1$ \footnote{Formally,\begin{equation*}
   \left. {\tan\xi=\frac{\sin(\bar{k}R-\ell\pi/2)\hat{f}(R)-\cos(\bar{k}R-\ell\pi/2)\hat{f}'(R)}{\cos(\bar{k}R-\ell\pi/2)\hat{f}(R)-\sin(\bar{k}R-\ell\pi/2)\hat{f}'(R)}}\right\rvert_{R\rightarrow\infty}.
\end{equation*} \protect\trick.} , and the amplitudes can be matched by scaling $\hat{f}_1(R)$ by a factor $C_1^{-1}$,
\begin{equation}
f_1(R)=C_1^{-1}\hat{f}_1(R).
\end{equation}
With both $\xi_1$ and $C_1$ fixed, $g_1(R)$ can be written as a linear combination of $\hat{f}_1(R)$ and $\hat{g}_1(R)$,
\begin{eqnarray}\label{eqn:qdt:matchphase}
g_1(R)=C_1\tan\lambda_1\hat{f}_1(R)+C_1\hat{g}_1(R),
\end{eqnarray}
as $\hat{f}_1$ and $\hat{g}_1$ span the solution space. Expressions for the QDT parameters can be found be considering Wronskians of appropriate pairs of reference functions (see appendix \ref{app:wronski})
We note that connecting $\{f_1(R),g_1(R)\}$ and $\{\hat{f}_1(R),\hat{g}_1(R)\}$ as in the above imposes a fixed relationship between
$\phi_1$ and $\xi_1$ at a particular energy, and that the QDT parameters $C_1$ and $\tan\lambda_1$ describing the transformation
\begin{equation}\label{eqn:qdt:wavefuncs}
  \begin{bmatrix} f_1 \\ g_1 \end{bmatrix} =
  \begin{bmatrix} C^{-1}_1 & 0 \\ C_1 \tan \lambda_1 & C_1 \end{bmatrix}
  \begin{bmatrix} \hat{f}_1 \\ \hat{g}_1 \end{bmatrix},
\end{equation}
will depend on the choice of these interrelated phases.

The above procedure for connecting up reference functions introduces $C_1$ and $\tan \lambda_1$ as the QDT parameters for the $i=1$ open entrance channel. To tie the QDT description to our physical system, we will (eventually) pick the pair of $\phi_1$ and $\xi_1$ such that $\xi_1$ reproduces the non-resonant scattering phase-shift $\delta_{\rm bg}$ [\textit{cf}. \eref{eqn:FeshbachPhase}]. For this choice, $c_2=0$ in \eref{sollong} which renders $f$ as the scattering wave-function in the non-resonant scenario. As such, $C^{-2}_1$ is the probability for the two atoms to penetrate to \blaa{short range} in absence of interchannel coupling, while $\lambda_1$ is the phase lag between $g_1$ and $\hat{g}_1$ due to the difference in kinetic energy at long range compared to \blaa{short range}. 
Together, the parameters $C_1$ and $\tan \lambda_1$ quantify the breakdown of the
WKB approximation [\textit{cf}. \eref{eqn:qdt:shortrange9i}] near threshold: at energies well above threshold
$C_1 \to 1$ and $\tan \lambda_1 \to 0$ as the WKB treatment becomes evermore valid at all separation ranges.
\subsection{Multichannel QDT}\label{sec:multichan}
In general, any open channel $i$ of a system can be subjected to the considerations for $i=1$ above.
The rationale behind defining $\hat f_i$ and $\hat g_i$ following the prescription of section \ref{sec:short} with a corresponding transformation
\begin{equation}\label{eqn:qdt:wavefuncs2}
  \begin{bmatrix} f_i \\ g_i \end{bmatrix} =
  \begin{bmatrix} C^{-1}_i & 0 \\ C_i \tan \lambda_i & C_i \end{bmatrix}
  \begin{bmatrix} \hat{f}_i \\ \hat{g}_i \end{bmatrix},
\end{equation}
becomes clear if we write the full complicated many-channel radial scattering wave function in terms of them.
\red{Suppose there is some interatomic distance $R_{\rm int}$ beyond which the channels
are essentially uncoupled and $V_1$ is well described by the van der Waals potential, \ie, the same condition for which \eref{eqn:qdt:long_range} emerged, but still at sufficiently short range such that all channels are locally open.
In this intermediate region the radial wave function matrix can be written in the form
\begin{equation}\label{eq:F}
\bm F(R) \sim \bm{\hat f}(R) + \bm{\hat g}(R) \bm Y .
\end{equation}}
Here $\bm{\hat f}$ and $\bm{\hat g}$ are diagonal matrices containing the $N=N_{\rm o}+N_{\rm c}$ channel reference function and $ \bm Y$ is a constant $N\times N$ matrix that plays a similar role to $\bm K_{\rm oo}$, but in this intermediate region where all channels are locally open.
As noted in section \ref{sec:short}$, \hat f_i$ and $\hat g_i$ only depend weakly on the
collision energy at \blaa{short range} since here $|\bar{E}_i| \ll |V_{i}^{\rm eff}(R)|$.
$\bm Y$ can therefore be considered constant with respect to the collision energy.
The energy dependence characteristic of the threshold behavior is instead captured by the QDT parameters, through the transformation \eref{eqn:qdt:wavefuncs2}.
As such, once $\bm Y$ is known, computing the physical scattering $\bm S$ matrix
at a given energy becomes simply a question of applying the appropriate
scattering boundary conditions, as detailed in Ref.~\cite{Mies1984} and Appendix \ref{sec:collapsY}.

In addition to the open entrance channel $\ket{1,1}\particleAdd\ket{1,1}$, the coupled equations in \eref{eqn:qdt:coupl} includes
closed channels (four in our case). In this intermediate region, these channels are locally open, so $\hat{f}$ and $\hat{g}$ are defined perfectly well following the prescription in section~\ref{sec:short}. Connecting to the classically forbidden region, where the wavefunction exponentially decays, introduces a single QDT parameter $\nu_i$ in each closed channel such
that
\begin{equation}\label{boundpar}
	\cos\nu_i \hat f_i - \sin\nu_i \hat g_i \overset{r \to \infty}\sim \frac{\ee^{-\lvert \bar{k}_i \rvert R}}{2\sqrt{\lvert \bar{k}_i \rvert}},
\end{equation}
where $\hat{f}_i$ and $\hat{g}_i$ are defined as in section \ref{sec:short}.
This identifies the particular linear combination of $\hat{f}_i$ and $\hat{g}_i$
which is decaying asymptotically. 

Given the QDT parameters in each channel, the asymptotic $\bm{S}$ matrix can be obtained
from the $\bm{Y}$ matrix by applying the appropriate scattering boundary
conditions~\cite{Mies1984}. We start from the $\bm Y$ matrix, which is split into sub blocks representing closed and open channels, 
\begin{equation}
  \bm Y  = \left[\begin{array}{c|c}
    \bm Y_\text{oo} & \bm Y_\text{oc}\\ \hline
    \bm Y_\text{co} & \bm Y_\text{cc}
  \end{array}\right].
\end{equation}
The procedure for eliminating the closed channels to connect ${\bm Y}$ to long-range, where the scattering boundary conditions for the $N_{\rm o}$ open channels are defined, is described in Appendix \ref{sec:collapsY}. Briefly, this relies on incorporating the effect of the closed channels on the open channels using the reduced $\bm Y$ matrix, $\bar{\bm Y}_\mathrm{oo}$:
\begin{equation}\label{eqn:y}
  \bar{\bm Y}_\mathrm{oo} = \bm Y_\mathrm{oo} - \bm Y_\mathrm{oc}\left(\tan \bm \nu  + \bm Y_\mathrm{cc}\right)^{-1}\bm Y_\mathrm{co},
\end{equation}
where $\tan\bm \nu$ is a diagonal matrix with $\tan \nu_i$ as diagonal elements. \Eref{eqn:y} therefore folds in the behavior of Feshbach resonances, which arise due to the closed channels and appear as poles when $\left|\tan \bm \nu  + \bm Y_\mathrm{cc}\right| \to 0$.
The effective reaction matrix, $\bar{\bm{R}}$ is given by [see \eref{eqn:app:Rbar}]
\begin{equation}\label{eqn:r}
  \bar{\bm  R} = \bm C^{-1}\left(\bar{\bm Y}_\mathrm{oo}^{-1} - \tan \bm \lambda\right)^{-1}\bm C^{-1},
\end{equation}
where $\bm C^{-1}$ and $\tan \bm \lambda$ are diagonal matrices containing elements $C_i^{-1}$ and $\tan\lambda_i$. \Eref{eqn:r} introduces the effects of scattering near threshold, such as a shape resonance.
Finally, the $\bm S$ matrix can be expressed as [see \eref{eqn:qdt:assymp}]
\begin{equation}\label{eqn:s}
  \bm S = \ee^{\ii \bm \xi}\left(1 + \ii\bar{\bm R}\right){\left(1 - \ii\bar{\bm R}\right)}^{-1} \ee^{\ii \bm \xi}.
\end{equation}
\subsection{Two Channel Model}
The general multichannel QDT framework outlined above in section \ref{sec:multichan} can be simplified in the case pertaining to a single open channel, and multiple closed channels over which a single isolated resonance resides. In this case an effective two channel QDT model captures the essential physics~\cite{mies2000, PhysRevA.88.052701}.  In our model, the open channel, o ($i=1$), is the $d$-wave entrance channel which contains the shape resonance, and the closed channel, c ($i=2$), contains the quasi-bound state giving rise to the Feshbach resonance. We choose the \blaa{short-range} reference functions such that the $\bm Y$ matrix is purely off-diagonal
\begin{equation}
  \bm Y =  \begin{bmatrix}0 & y \\ y & 0 \end{bmatrix},
\end{equation}
which we are always free to do in the two channel case~\cite{giusti-suzor.fano:alternative}.

Applying the above MQDT formulae \eref{eqn:y} and \eref{eqn:r} gives 
\begin{equation}
  \bar{\bm R} = \frac{y^2C^{-2}}{-\tan\nu - y^2\tan\lambda},
\end{equation}
where in our notation we suppress the $i=1,2$ indices of the of the QDT parameters.
Writing the $\bm S$ matrix---in our case just a single complex number---in terms of the scattering phase shift,
$S = \ee^{2\ii\delta_{d}}$, and using \eref{eqn:s} gives~\footnote{We note the trigonometric identity
\begin{equation*}
  \arctan z = -\frac{\ii}{2}\ln\left(\frac{1 + \ii z}{1 - \ii z}\right).
\end{equation*} \protect\trick.
}
\begin{equation}
  \delta_{d} = \xi + \arctan{\bar{\bm R}},
\end{equation}
which yields the scattering phase shift in terms of the QDT parameters:
\begin{equation}
  \delta_{d} = \xi + \arctan{\left(\frac{y^2C^{-2}}{-\tan\nu - y^2\tan\lambda}\right)}.
\end{equation}
Using a linear expansion of $\tan\nu$ which goes to zero in the vicinity of a resonance
$\tan\nu \approx \left.\frac{\partial\nu}{\partial E}\right|_{E = E_0}(E - E_0)$
and defining $\frac{1}{2}\bar\Gamma = \left(\left.\frac{\partial\nu}{\partial E}\right|_{E = E_0}\right)^{-1}y^2$ gives
\begin{equation}\label{eqn:mqdt}
  \delta_{d} = \xi + \arctan{\left(\frac{\frac{1}{2}\bar\Gamma C^{-2}}{E_0 - E - \frac{1}{2}\bar\Gamma\tan\lambda}\right)}.
\end{equation}
Comparing the above to \eref{eqn:FeshbachPhase}, we note that $\xi$ corresponds to our measured background phase shift, $\delta_\mathrm{bg}$.

We now consider the resonance position as a function of the external magnetic
field. In the closed channel, the bare bound state position in energy is $E_0 = \delta\mu(B - B_0)$ where $\delta\mu$ is the difference in magnetic moment between the open and closed channels, and
$B_0$ is the field at which the (non-interacting) resonance is at
threshold~\cite{RevModPhys.82.1225}. By defining $\bar{\Gamma}_B = \bar{\Gamma}/\delta\mu$,
\begin{equation}\label{eqn:mqdt:B}
  \delta_{d} = \delta_\mathrm{bg} + \arctan{\left(\frac{\frac{1}{2}\bar\Gamma_{B} C^{-2}}{B - (B_0 + E/\delta\mu + \frac{1}{2}\bar\Gamma_{B}\tan\lambda)}\right)}.
\end{equation}
Within this model, the width and position of the resonance in magnetic field are therefore given by
\begin{equation}
  \label{eqn:qdt:Bwidth} \Gamma_B(E) = C^{-2}(E) \bar{\Gamma}_B,
  \end{equation} and
  \begin{equation}
  \label{eqn:qdt:Bres} B_\text{res}(E) = B_0 + \frac{E}{\delta\mu} + \frac{\bar{\Gamma}_B}{2} \tan\lambda(E),
\end{equation}
respectively.
These formulae elegantly demonstrate the advantage of the MQDT approach.
The width of the resonance is factorised into one energy-dependent part associated with the long-range threshold effects, $C^{-2}(E)$, and another energy-independent part, $\bar{\Gamma}_B$, associated with the short-range physics. They also demonstrate that threshold effects modify not only the width of a resonance but also its
position~\cite{mies.raoult:analysis,PhysRevA.70.012710,PhysRevA.100.042710}.
Within Fano's configuration-interaction approach the shift in the resonance
position is due to the off-energy shell interactions~\cite{PhysRev.124.1866},
which have the effect of mixing in the irregular solution to the bare
scattering solution in the open channel.
Having chosen the reference function $f_1$ to have a phase that matches the physical background scattering
phase shift $\delta_\mathrm{bg}$ in the $d$-wave channel (\ie{} to be the regular solution), the admixture of the irregular solution $g$ is completely captured by $\tan\lambda$ within the QDT formalism, which determines the shift~\cite{fano:connection}.
\subsection{Computations}
We now consider the practical computation of the QDT parameters. 
These can be computed either analytically~\cite{gao:solutions} or
numerically~\cite{yoo.greene:implementation,PhysRevA.84.042703,ruzic.greene.ea:quantum}. 
Here we implement the stable numerical approach developed by Ruzic~\etal~\cite{ruzic.greene.ea:quantum}.
As discussed by Ruzic~\etal, reference solutions lose their linear
independence when propagating through a centrifugal barrier, so it is optimal to choose reference functions which are purely exponentially growing and decaying in that region. 
In our case this simply (by construction) corresponds to choosing $\phi_1 = 0$ in \eref{eqn:qdt:WKBPhase}, as can be seen in \figref{fig:wavefunction}.
We then combine this approach with a reference-function rotation to obtain any particular set of reference functions~\cite{giusti-suzor.fano:alternative,osseni.dulieu.ea:optimization, PhysRevA.86.022711}. 
This rotation produces QDT parameters corresponding to a different $\phi_1$ such that $\xi = \delta_\mathrm{bg}$ as discussed earlier.
\begin{figure}
    \centering
    \includegraphics[width=\linewidth]{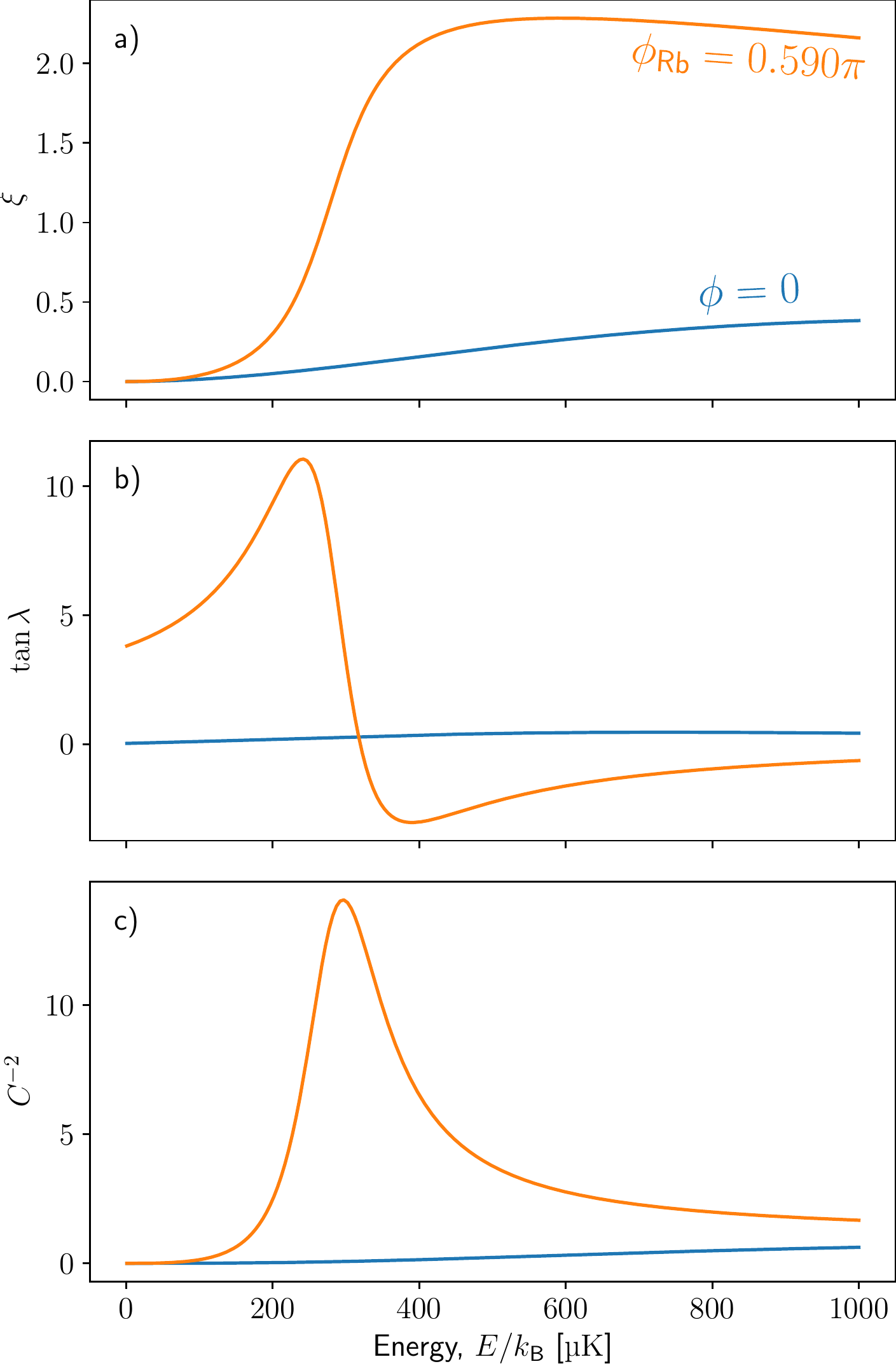}
    \caption{QDT parameters, $\xi$ (a), $\tan\lambda$ (b), and $C^{-2}$ (c) as a function of energy. The blue lines show numerically calculated using QDT reference functions based on phase $\phi_1 = 0$ [\textit{cf.} \eref{eqn:qdt:WKBPhase}]. The orange lines show the QDT parameters transformed to $\phi_1=0.590\times\pi$ using the analytic rotation formulae Eqns.~\ref{eqn:rotation}.}
    \label{fig:qdt_rotate}
\end{figure}
The QDT parameters are computed in the same way as detailed in Ref.~\cite{Mies1984}, and here we just highlight details specific to this work.
Numerical propagation of the reference functions was done using Numerov's method~\cite{Numerov}. 
The $\hat{f}_1$ reference function was propagated out from $R_\mathrm{min} = 0.1$ to $R_\mathrm{max} = 25$ using \eref{eqn:qdt:shortrange9} as the short-range boundary condition with $\phi_1 = 0$.
Matching $\hat{f}_1$ with $f_1$ defines $\xi_1$ via \eref{eqn:qdt:long_range6} \cite{Note3}. 
This also defines the reference function $g$ via \eref{eqn:qdt:long_range7}, which serves as a boundary condition to propagate $g_1$ back to short range. The QDT parameters can be extracted using Eqns.~(\ref{eq:wron1}) and (\ref{eq:wron2}). These QDT parameters are then rotated~\cite{giusti-suzor.fano:alternative,osseni.dulieu.ea:optimization, PhysRevA.86.022711} such that $\xi = \delta_\mathrm{bg}$ following the procedure in Appendix~\ref{app:rotate}.
We note that because $\phi_i$ is defined at short range (where both the collision energy and the centrifugal term are small compared to the depth of the potential) a single energy independent $\phi_1$ will reproduce the energy dependent $\delta_\mathrm{bg}$ over the entire range of energies we are interested in here.

\Figref{fig:qdt_rotate} shows the QDT parameters obtained, both as propagated using $\phi_1=0$, and following an analytic rotation so that $\xi$ matches the experimentally observed background scattering phase-shift.
The choice of $\phi_1 = 0$ produces slowly varying QDT parameters, while those rotated to match the physical scattering phase shift show a peak in $C^{-2}$ which directly gives the increased probability of tunneling through the $d$-wave barrier at the energy of the observed shape resonance.

The QDT parameters are not sensitive to the choice of propagation limits: it is sufficient that at $R_\mathrm{min}$ WKB is valid, and at $R_\mathrm{max}$ the potential has decayed to near-zero. The propagation of the reference functions relies only on knowing the reduced mass $\mu$, the van der Waals coefficient $C_6$, and the angular momentum $\ell$. This means that the QDT parameters are solely a property of the general long-range potential and account for the threshold effects on the scattering.
\begin{figure}
  \centering
  \includegraphics[width=\linewidth]{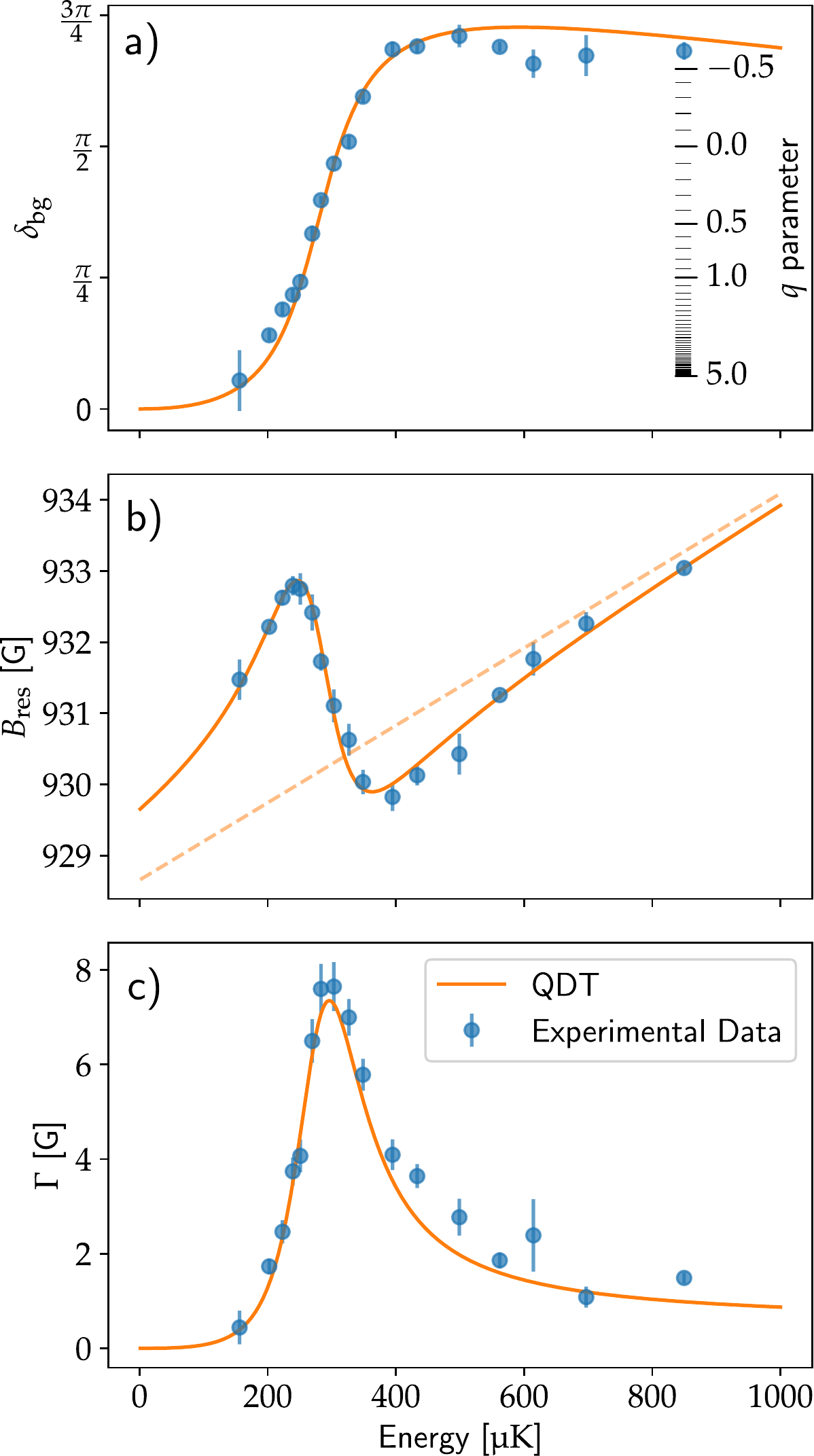}
  \caption{\label{fig:res_data} The properties of the Feshbach
    resonance as measured at a number of collision energies: {a)} the
    background phase shift and corresponding $q$ parameters, {b)} the magnetic field of the
    resonance, and {c)} the
    width of the resonance. We include predictions from the multichannel quantum defect theory model. The dashed line in b) indicates the expected position of the Feshbach resonance in absence of the open-channel effects (shape resonance).
 }
\end{figure}
\section{Results}
For a range of energies, we measure of the scattering fraction as a function of magnetic field to obtain scans as those shown in \figref{fig:fano}. From these magnetic field scans we extract the resonance parameters shown in \figref{fig:res_data}, namely the background phase $\delta_\text{bg}$, the resonance position $B_\text{res}$, and the resonance width $\Gamma_B$, as defined by \eref{eqn:FeshbachPhase}. These three parameters completely describe the observed resonance features and their variation captures the interplay between the two resonances.
\subsection{Experimental Observations}\label{sec:results}
\Figref{fig:res_data}a shows the increase of the open channel $d$-wave background phase shift
across the shape resonance. 
In particular we note its transition through the value $\pi/2$ at the location of the
shape resonance. 
During the course of this, the Fano profile undergoes a $q$-reversal which
flips the shape of the Fano profiles shown in \figref{fig:fano}. The
background phase changes by a total value of less than $3\pi/4$ in
this system, while an isolated resonance normally accrues a total phase
change of $\pi$ asymptotically---a general feature of resonances in both quantum and classical systems. 
The discrepancy can be explained by considering that the shape resonance is not a pure,
isolated Breit-Wigner resonance: not only are there other resonances in
the channel, but in the in the absence of the shape resonance the background phase-shift of the channel would increase~\cite{PhysRevA.70.012710,Sadeghpour2000}.

\Figref{fig:res_data}b displays the magnetic field at which the resonance feature is positioned, where we observe a `kink' in the trajectory, shifting by a substantial fraction of the
width of the resonance.  Above threshold, a Feshbach resonance usually moves linearly in energy as shown by the dashed line, with the slope given by the difference in magnetic moment between the two channels. The deviation from linear is the manifestation of the interaction between states. Indeed, examples of such behaviour has previously been found  for a Feshbach resonance  interacting with an antibound state
\cite{Marcelis2004,Thomas2018}, and a $p$-wave shape resonance\cite{ahmedbraun2021probing}.

As shown in \figref{fig:res_data}c, the Fano profile broadens across the nominal shape resonance energy position by orders of magnitude from the zero-energy width. The Feshbach resonance we inspect is considered narrow~\cite{PhysRevLett.89.283202}, and at zero-energy the width is limited by the weak $s$ to $d$-wave coupling, and its observation hence requires a very stable and low noise magnetic field. For experiments conducted above threshold, the Feshbach resonance is, however, readily detected through the shape resonance.
\begin{figure}
  \centering
  \includegraphics[width=\linewidth]{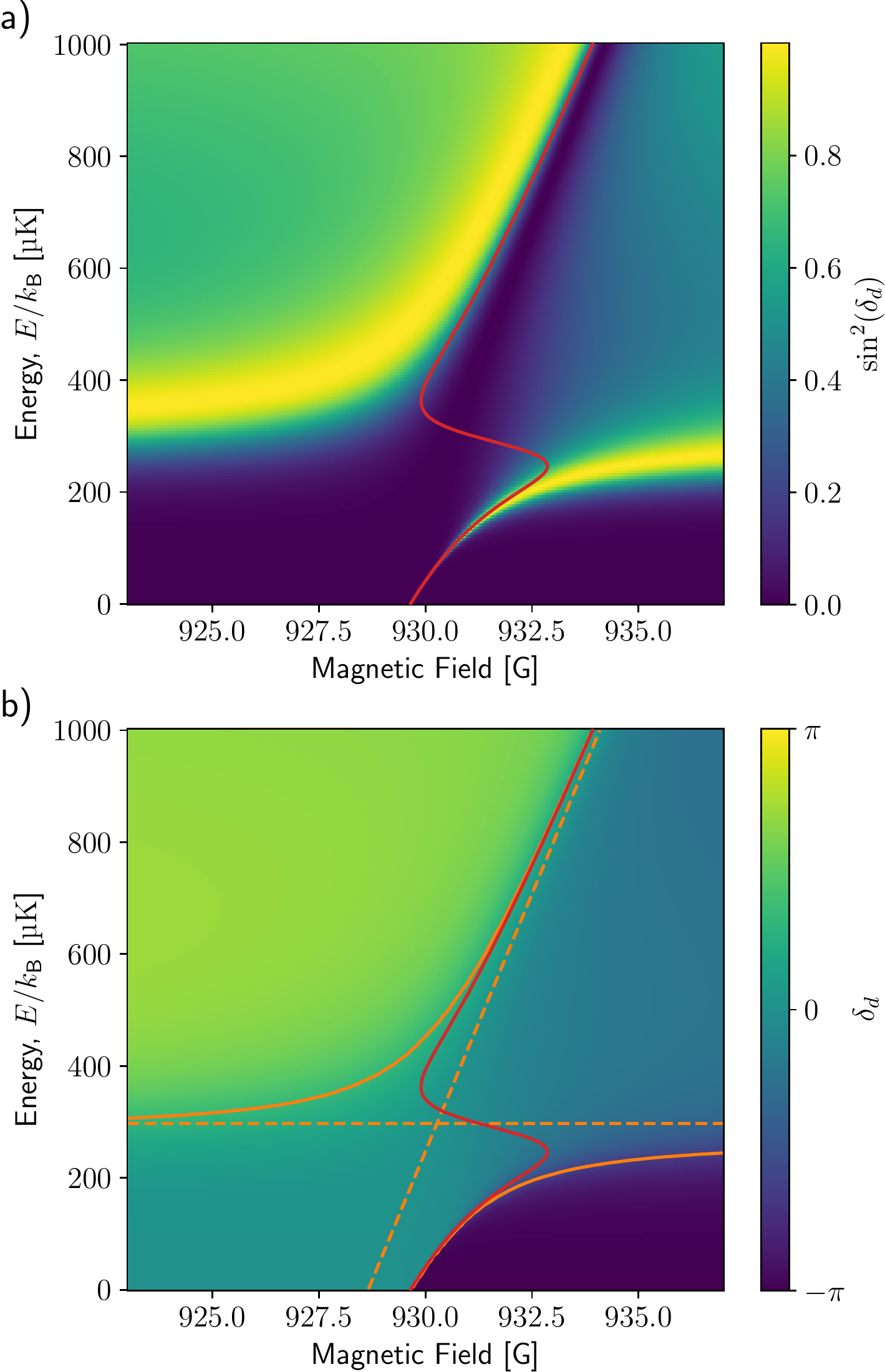}
  \caption{\label{fig:res_centre}MQDT calculations in both collision energy and magnetic field. a) $\sin^2\delta_d$, which is proportional to the scattering cross section. b) the scattering phase shift. On both panels, the position of the Fano resonance in magnetic field is shown in red, while in b), the resonance positions in energy are shown as solid orange lines, and the non-interacting resonance positions are shown as dashed orange lines.}
\end{figure}
\subsection{QDT Analysis}
We find the short-range QDT phase $\phi = 0.590\pi$ in the open channel by fitting $\xi$ to the observed background phase in \figref{fig:res_data}a. The QDT parameters corresponding to this $\phi$ are shown in \figref{fig:qdt_rotate}. By fitting \eref{eqn:qdt:Bres} to $B_\text{res}$ (\figref{fig:res_data}b), we obtain the Feshbach resonance parameters $\bar{\Gamma} = \SI{96}{\micro\kelvin}$, $\delta\mu =
\SI{184}{\micro\kelvin\per\gauss}$, and $B_0 = \SI{928.7}{\gauss}$. 

The width of the resonance predicted by \eref{eqn:qdt:Bwidth} is shown as an orange line in \figref{fig:res_data}c. This is in excellent agreement with the experimental observations and describes the energy dependence of the width entirely through $C^{-2}$. Since $C^{-2}$ quantifies the tunnelling through the centrifugal barrier to short range, we can attribute the broadening of the resonance to the increased amplitude of the wavefunction at short range due to the shape resonance.

The shift in $B_\text{res}$ due to the interaction with the open channel is given by the last term of \eref{eqn:qdt:Bres},
\begin{equation} \label{eqn:qdt:Bshift}
  \delta B = -\frac{\bar{\Gamma}_B}{2} \tan\lambda.
\end{equation}
The calculated $\tan\lambda$ for this system, shown in \figref{fig:qdt_rotate}b,
explains the non-linear and non-monotone resonance trajectory. We observe that $\tan\lambda$ is non-zero at threshold so the zero-energy position of the
Feshbach resonance is already shifted by $\sim\SI{1}{\gauss}$ due to the coupling to
the open channel, that is, due to the presence of the shape resonance. This is particularly apparent in \figref{fig:res_data}b which shows the uncoupled resonance position $B_\text{0} + E /\delta\mu$ as a dashed line.

\Figref{fig:res_centre}a shows the sine-squared of the scattering phase shift predicted by the MQDT model, which is proportional to the scattering cross-section. The red line shows the predicted position of the Feshbach resonance in magnetic field (the orange line in \figref{fig:res_data}b), which can be seen to move between two regions of strong scattering as the energy increases. Resonances are associated with a rapid change in the scattering phase by $\pi$. In \figref{fig:res_centre}b we present the scattering phase with the positions of the resonances in both the energy and field. We locate resonances in energy by the position at which the change in phase with energy is maximal, \ie,
\begin{equation}
    \frac{\partial^2 \delta}{\partial E^2}\Big\rvert_B = 0.
\end{equation}
These maxima correspond to positions where the phase winds by $\sim \pi$, which is characteristic of a resonance. Similarly, $B_\text{res}$ corresponds to a winding of $\pi$ in field. As is clear from the figure, the positions of the resonances in energy and field do not always line up. 
In the middle of the kink ($\sim \SI{300}{\micro\kelvin}$), one encounters a magnetic resonance where there is no resonance in energy. As previously discussed, (energy) resonances arise due to the coupling with a quasi-bound state near the collision energy. Here however, one can see that the Fano profile (a dip) arises not from a nearby quasi-bound state, but from the temporary absence of one. This is discussed in our previous work \cite{Chilcott2021A}, where we have shown that this occurs as the quasi-bound states associated with the two resonances undergo an avoided crossing. 
The lack of correspondence in energy and field positions is also clear from equations~\eqref{eqn:FeshbachPhase} and \eqref{eqn:mqdt:B}: only $B$ in the denominator changes as a function of magnetic field, giving rise to an isolated Fano profile; in energy, $\Gamma_B$, $B_\text{res}$ and $\delta_\text{bg}$ all vary rapidly across the shape resonance, leading to a non-trivial winding of the scattering phase.
Raoult and Mies~\cite{PhysRevA.70.012710} state this another way: one cannot always assign a meaningful energy width to a Feshbach resonance due to the energy shift. Here, we see the energy shift effectively splits the resonance in two. However, we observe that the Fano profile in magnetic field is always singular and well defined: the width of the resonance is clear.
\section{Conclusion}
In this work we have studied the non-trivial interplay between a shape resonance and a Feshbach resonance in ultracold atomic \Rb{} collisions.
By manipulating the collision energy and magnetic field we can tune the shape parameter, $q$, of the Fano profile over a range sufficient to observe a full $q$-reversal.
In addition to the $q$-reversal we observe strong broadening, and an oscillatory kink in the resonance trajectory as the Feshbach resonance moves over the shape resonance.

To explain this behaviour, we have presented a multichannel quantum defect theory (MQDT) analysis of the experimental data.
The MQDT model is able to accurately capture the essential physics of the interactions over the entire range of energy and magnetic field of interest in terms of just 4 constants ($\phi = 0.590\pi$, $\bar{\Gamma} = \SI{96}{\micro\kelvin}$, $\delta\mu = \SI{184}{\micro\kelvin\per\gauss}$, and $B_0 = \SI{928.7}{\gauss}$) and the three energy dependent QDT parameters ($C^{-2}$, $\tan\lambda$, and $\xi$), which are simply properties of the long range van der Waals potential. 

We observe an excellent match between experiment and theoretical predictions, and the MQDT framework demonstrates that the observed resonance behaviour is primarily due to the open-channel, related to the short-range enhancement (determined by $C^{-2}$) and long-range phase rotation (determined by $\tan\lambda$) of the scattering wave-function.
In addition to providing additional insight MQDT also proves a vastly simpler tool than complete coupled-channels calculations which require a complex multichannel potential.

Our experimental scheme using an optical collider implements a Feshbach resonance ``microscope'' which magnifies a narrow zero-energy feature through a shape resonance. Threshold behaviour dictates that an isolated Feshbach resonance will generally broaden as its position is tuned towards higher energies with a magnetic field \cite{Horvath_2017}. The shape resonance expedites this broadening while maximising the number of scattered particles and the signal to noise ratio for the measurement. While for our particular realization, the Feshbach resonance in question can be observed close to threshold, in the future the approach may be used to verify predicted ultra-narrow Feshbach resonances that evades experimental observation in conventional loss spectroscopy.
\begin{acknowledgments}
This work was supported by the Marsden Fund of New Zealand (Contract No. UOO1923). J. F. E. Croft acknowledges a Dodd-Walls Fellowship and M. Chilcott a
University of Otago Postgraduate Publishing Bursary (Doctoral).
\end{acknowledgments}


%

\appendix

\section{Loss Spectroscopy}\label{app:loss}

\begin{figure}
  \centering
  \includegraphics[width=\linewidth]{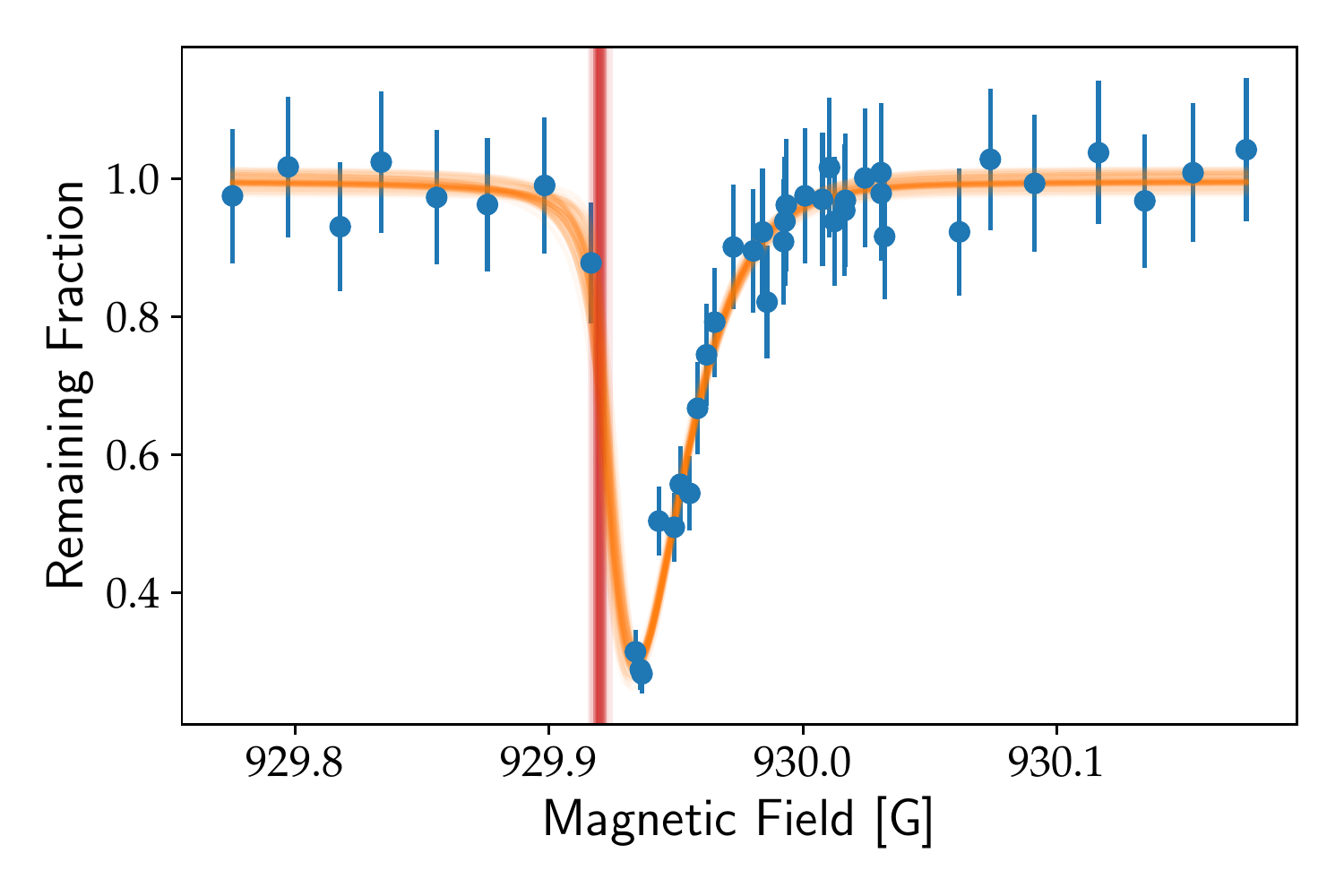}
  \caption{\label{fig:loss}A representative measurement of the atom
    loss fitted by the described model. The extracted resonance
    position is shown in red.}
\end{figure}

We measure the (near) zero-energy position of the Feshbach resonance by observing the effect of a magnetic field on a stationary atomic cloud. At the Feshbach resonance, the scattering length between the \Rb{} atoms diverges, resulting in an increased three-body loss rate.

The procedure for performing loss spectroscopy measurement is initially identical to that laid out in Section~\ref{sec:experimental}, up to the point where we would split and collide the clouds of atoms. Instead, a single cloud which has been further evaporatively cooled below \SI{400}{\nano\kelvin} is held in a stationary optical dipole trap and exposed to a magnetic field for \SI{200}{\milli\second}).  An absorption image of the single cloud is then used to estimate the number of atoms remaining in the trap. In such an experiment, the profile of the atom loss can often be well approximated by a Gaussian line shape~\cite{PhysRevLett.89.283202}, especially when the width of the Feshbach resonance is greater than the range of collision energies present in a thermal cloud. In our data, we observe an asymmetry that is distinctly non-Gaussian (shown in \figref{fig:loss}), requiring us to take into account the thermal distribution of the finite temperature cloud---the Maxwell-Boltzmann distribution is skewed towards high energies with negative energies forbidden and the same skew is imposed upon the shape of the atom loss in magnetic field as the resonance tunes above threshold.
Explicitly, we model the coupling rate to the Feshbach state as a Breit-Wigner profile in energy,
\begin{equation}
  R(E, B) \propto \frac{\gamma}{\gamma^2 + (E - \epsilon(B))^2},
\end{equation}
for atoms at a given energy $E$, when the Feshbach resonance is at energy $\epsilon(B) = \delta\mu(B - B_0)$. The distribution of kinetic energies (which are strictly positive) are taken to be Maxwellian at a temperature $T$:
\begin{equation}
  P(T,E) = kT e^{-E/kT}.
\end{equation}
The three-body loss rate $K_3$ is then proportional to the integral of these over energy,
\begin{equation}
  K_3(B) \propto \int_0^\infty R(E, B) P(T, E) \,\text{d} E.
\end{equation}
If we assume that the loss process does not produce evaporative
heating/cooling, and that the loss rate from other processes is negligible, then the loss can be modelled by
\begin{equation}
  \dot{N}(B) = - K_3(B) N^2,
\end{equation}
where $N$ is the number of atoms remaining. Here we are describing the
three-body loss process as second-order in atom number (akin to a two-body loss process) to encapsulate the inverse density and hence $N^{-1}$ atom number dependence of $K_3$~\cite{Beaufils2009}. A fit of this model to experimental data is shown in \figref{fig:loss}, and a number of such measurements give us an estimate of the zero-energy Feshbach resonance position $B_0 = \SI{929.918(6)}{\gauss}$.

\section{QDT Supplementary}
\subsection{Calculating QDT parameters}\label{app:wronski}
Abel's identity implies that the Wronskian $W$ of any pair of solutions to \eref{eqn:qdt:long_range2} is a constant independent of $R$. By considering  Wronskians of appropriate combinations of the QDT reference funtions $\{f_i,g_i,\hat{f}_i,\hat{g}_i\}$ expressions for the QDT parameters can be obtained \cite{Mies1984,ruzic.greene.ea:quantum}. For example, from \eref{eqn:qdt:wavefuncs2}, and $W(\hat{g}_i,\hat{f}_i)=1$ and $W(\hat{g}_i,\hat{g}_i)=0$  
\begin{subequations}\label{eqn:qdt:parameters}
\begin{align}
  W(\hat{g}_i,f_i)& =C_i^{-1}  ,\\
 W(\hat{g}_i,g_i)&= C_i  \tan \lambda_i,
\end{align}
\end{subequations}
and evaluations around $R \simeq R_m$ and in the limit $R\rightarrow\infty$ give
\begin{subequations}\label{eq:wron}
\begin{eqnarray}
	 C_i^{-2} &=& [W(\hat{g}_i,f_i)]^2\nonumber
	 \\&=& \begin{cases}
\bar{k}_i(R) f_i^2(R)  + f_i'^2(R)/\bar{k}_i(R),\text{for }R \simeq R_m\\
	[\bar{k}_i \hat{f}_i^2(R)  + \hat{f}_i'^2(R)/\bar{k}_i]^{-1},\text{for } R\rightarrow\infty
	\end{cases},\phantom{hhsdf}\label{eq:wron1}
\end{eqnarray}
\begin{eqnarray}
\tan\lambda_i &=& C_i^{-1}W(\hat{g}_i,g_i)\nonumber
	 \\&=& \begin{cases}\begin{aligned}[b]
&\bar{k}_i(R) f_i(R)g_i(R)\\&+f'(R)_1g'(R)_i/\bar{k}_i(R) \end{aligned},\text{for }R \simeq R_m\\
	\begin{aligned}[b]-C_i^{-2}[&\bar{k}_i \hat{f}_i(R)\hat{g}_i(R)\\&+\hat{f}'_i(R)\hat{g}'_i(R)/\bar{k}_i]\end{aligned},\text{for } R\rightarrow\infty
	\end{cases}.\phantom{hhsdf}\label{eq:wron2}
\end{eqnarray}
\end{subequations}
\subsection{Open channel elimination in MQDT and expression of the  \texorpdfstring{$\bm S$}{S} matrix}\label{sec:collapsY}
In this section, we relate the constant $N\times N$ $\vec{Y}$ matrix introduced in \eref{eq:F} to the $N_{\rm o}\times N_{\rm o}$ $\bm S$ matrix~\cite{Mies1984,ruzic2015exploring}.

As discussed in section \ref{sec:multichan}, a solution matrix to the coupled channels problem over some (intermediate) short range can be expressed through the QDT reference functions $\hat f_i$ and $\hat g_i$ as
\begin{equation}\label{eq:App}
    {\bm F} = \hat{\bm f} + \hat{\bm g}{\bm Y},
\end{equation}
where $\bm Y$ is constant matrix.
This is possible because at this intermediate range (\textit{cf}. section~\ref{sec:multichan}), where the boundary conditions define the $N\times N$ diagonal matrices $\hat{\bm f}$ and $\hat{\bm g}$, all $N$ channels are locally open---even the $N_{\rm c}$ channels that are asymptotically closed.

To obtain the $N_{\rm o}$ physically meaningful solutions, the closed channels need to be eliminated from \eref{eq:App}. This elimination can be done by considering a transformation ${\bm T}$ that builds a reduced $N\times N_{\rm o}$ solution matrix $ \bar{\bm F}$ out of ${\bm F}$, where each of $N_{\rm o}$ column of $ \bar{\bm F}$ of is linear combination of the $N$ columns of ${\bm F}$
\begin{eqnarray}
\overbrace{   \begin{bmatrix}
        \bar{\bm F}_\text{oo}\\
        \bar{\bm F}_\text{co}
    \end{bmatrix}}^{\bar{\bm F}}&=&\overbrace{  
\left(
    \begin{bmatrix}
    \hat{\bm f}_\text{oo} & 0\\
    0 & \hat{\bm f}_\text{cc}
    \end{bmatrix} 
    +
    \begin{bmatrix}
    \hat{\bm g}_\text{oo} & 0\\
    0 & \hat{\bm g}_\text{cc}
    \end{bmatrix}
    \begin{bmatrix}
    {\bm Y}_\text{oo} & {\bm Y}_\text{oc}\\
    {\bm Y}_\text{co} & {\bm Y}_\text{cc}
    \end{bmatrix}
    \right)}^{\bm F}\overbrace{
    \begin{bmatrix}
        {\bm T}_\text{oo}\\
        {\bm T}_\text{co}
    \end{bmatrix}}^{\bm T}\nonumber\\
   & =&\begin{bmatrix}
    \hat{\bm f}_\text{oo} {\bm T}_\text{oo} +  \hat{\bm g}_\text{oo} ({\bm Y}_\text{oo} {\bm T}_\text{oo}+{\bm Y}_\text{oc} {\bm T}_\text{co}) \\
     \hat{\bm f}_\text{cc}{\bm T}_\text{co}  + \hat{\bm g}_\text{cc}({\bm Y}_\text{co}{\bm T}_\text{oo}+{\bm Y}_\text{cc}{\bm T}_\text{co})
    \end{bmatrix} \nonumber\\
     & =&\begin{bmatrix}
    \hat{\bm f}_\text{oo}  +  \hat{\bm g}_\text{oo}  \bar{\bm Y}_\text{oo} \\-
     ({\underbrace{\hat{\bm f}_\text{cc} - \tan {\bm \nu}_\text{cc} \hat{\bm g}_\text{cc}}_\text{$\underset{R \to \infty}\sim 0$, \textit{cf.} \eref{boundpar}}}) (\tan {\bm \nu}_\text{cc}+{\bm Y}_\text{cc})^{-1}{\bm Y}_\text{co}
    \end{bmatrix},\phantom{aksk}
\end{eqnarray}
where in order to obtain the last step the blocks of ${\bm T}$ are chosen as ${\bm T}_\text{oo}=1$ and ${\bm T}_\text{co}=-(\tan {\bm \nu}_\text{cc}+{\bm Y}_\text{cc})^{-1}{\bm Y}_\text{co}$, and we introduced the 
reduced ${\bm Y}$ matrix, $\bar{\bm Y}_\text{oo}={\bm Y}_\text{oo} {\bm T}_\text{oo}+{\bm Y}_\text{oc} {\bm T}_\text{co}$, which can be cast as \eref{eqn:y}. The blocks of the resulting reduced solution matrix fulfill
\begin{subequations}\label{eqn:ybar:Fasymp}
\begin{align}
    \bar{\bm F}_\text{oo} & = \hat{\bm f}_\text{oo} + \hat{\bm g}_\text{oo}\bar{\bm Y}_\text{oo},\label{eqn:ybar:Fasymp:open}\\
    \bar{\bm F}_\text{co} & \underset{R \to \infty}\sim 0,
\end{align}
\end{subequations}
and the closed channels have been eliminated.
\Eref{eqn:ybar:Fasymp:open} expresses the open-open block of the reduced solution matrix $\bar{\bm F}_\mathrm{oo}$ in terms of the \blaa{short-range} reference functions $\hat f_i$ and $\hat g_i$. However, to make the connection to the physical $\bm S$ matrix we want a solution of the form
\begin{equation}\label{eqn:R_definition}
    {\bm F}_\mathrm{oo} \sim {\bm f} + {\bm g} \bar{\bm R},
\end{equation}
based on the energy normalized asymptotic reference functions $f_i$ and $g_i$.
$\bar{\bm R}$ constitutes an effective reaction matrix with
the `true' reaction matrix (often called $K$ or $R$ in the literature) being defined by a form identical to \eref{eqn:R_definition}, but with the QDT reference solutions $f_i$ and $g_i$ replaced with the appropriate spherical Bessel function solutions [\textit{cf.} \eref{eqn:cc1}].

The relation between the short- and long-range reference function is [\eref{eqn:qdt:wavefuncs2}]
\begin{subequations}\label{eq:appfcomb}
\begin{align}
    \hat{\bm f}_\text{oo} & = {\bm C}{\bm f}_\text{oo}, \label{eq:appf1}\\ 
    \hat{\bm g}_\text{oo} & = {\bm C}^{-1}{\bm g}_\text{oo} - \tan {\bm \lambda \bm C \bm f}_\text{oo},\label{eq:appf12}
\end{align}
\end{subequations}
and inserting Eqns.~\ref{eq:appfcomb} into \eref{eqn:ybar:Fasymp:open} gives
\begin{align}
    \bar{\bm F}_\text{oo} &  = {\bm C} {\bm f}_\text{oo} + ({\bm C}^{-1} {\bm g}_\text{oo} - \tan{\bm \lambda}{\bm C}{\bm f}_\text{oo})\bar{\bm Y}_\text{oo} \nonumber\\
     & = {\bm f}_\text{oo}{\bm C}(1-\tan{\bm \lambda}\bar{\bm Y}_\text{oo})+{\bm g}_\text{oo}{\bm C}^{-1}\bar{\bm Y}_\text{oo}
\end{align}
Multiplying from the right by ${\bm C}(1-\tan{\bm \lambda}\bar{\bm Y}_\text{oo})^{-1}$ a transformed solution matrix of the desired form \eref{eqn:R_definition} is obtained:
\begin{align}\label{eqn:app:Rbar}
    \bar{\bar{\bm F}}_\text{oo} &  = {\bm f}_\text{oo} + {\bm g}_\text{oo}\underbrace{{\bm C}^{-1}(\bar{\bm Y}_\text{oo}^{-1} - \tan{\bm \lambda})^{-1}{\bm C}^{-1}}_{\bar{\bm R}}.
\end{align}
Using the asymptotic properties of $f_i$ and $g_i$ [\textit{cf.} Eqns.~\ref{eqn:qdt:longrange}], this solution can be expanded as
\begin{eqnarray}\label{eqn:app:Rbar2}
\bar{\bar{\bm F}}_\text{oo}&  \overset{R \to \infty}\sim&
\frac{\ee^{\ii ({\bm k}R - {\bfell}\frac{\pi}{2}+{\bm \xi})}-\ee^{-\ii ({\bm k}R - {\bfell}\frac{\pi}{2}+{\bm \xi})}}{2\ii}
 \nonumber\\ & & +\frac{\ee^{\ii ({\bm k}R - {\bfell}\frac{\pi}{2}+{\bm \xi})}+\ee^{-\ii ({\bm k}R - {\bfell}\frac{\pi}{2}+{\bm \xi})}}{2}\bar{\bm R}\nonumber\\
&\sim&\ee^{-\ii ({\bm k}R - {\bfell}\frac{\pi}{2})}\ee^{-\ii {\bm \xi}}(1 -\ii\bar{\bm R})+\ee^{\ii ({\bm k}R - {\bfell}\frac{\pi}{2})}\ee^{\ii {\bm \xi}}(1+ \ii\bar{\bm R}),\nonumber\\
\end{eqnarray}
and by multiplying from the right by $(1 -\ii\bar{\bm R})^{-1}\ee^{\ii {\bm \xi}}$, a solution form with incoming and outgoing spherical wave components is obtained:
\begin{equation}\label{eqn:qdt:assymp}
\bar{\bar{\bar{\bm F}}}_\text{oo} \overset{R \to \infty}\sim\underbrace{\ee^{-\ii ({\bm k}R - {\bfell}\frac{\pi}{2})}}_\text{incoming}+\underbrace{\ee^{\ii ({\bm k}R - {\bfell}\frac{\pi}{2})}}_\text{outgoing}\overbrace{\ee^{\ii {\bm \xi}}(1+ \ii\bar{\bm R})(1- \ii\bar{\bm R})^{-1}\ee^{\ii {\bm \xi}}}^{{\bm S}}.
\end{equation}
In particular, it provides us with the desired expression of ${\bm S}$, \eref{eqn:s}.
\subsection{Rotation of QDT parameters}\label{app:rotate}
Once the QDT parameters have been calculated for $\phi = 0$, it is straightforward to analytically obtain them for any particular choice of $\phi$ by using the transformations ~\cite{PhysRevA.86.022711}:
\begin{subequations}\label{eqn:rotation}
\begin{equation}
  \bar{\xi} = \arctan\left[\frac{C^2 \sin\xi (\cos\phi + \tan\lambda \sin\phi) - \cos\xi\sin\phi}{C^2\cos\xi(\cos\phi + \tan\lambda\sin\phi) + \sin\xi\sin\phi}\right],
\end{equation}
\onecolumngrid
\begin{equation}
  \tan\bar{\lambda} = - \frac{2C^4\tan\lambda\cos 2\phi + \left[1 + C^4(\tan^2\lambda - 1)\right]\sin 2\phi}{2C^4\cos^2\phi + 2\sin\phi\left[\sin\phi + C^4\tan\lambda(2\cos\phi + \tan\lambda\sin\phi\right]},
\end{equation}
\begin{equation}
  \bar{C} = \left[\frac{\sin\xi\sin\phi}{C} + C\cos\xi(\cos\phi +\tan\lambda\sin\phi)\right]
  \times \sqrt{1 + \frac{\left[\cos\xi\sin\phi - C^2\sin\xi(\cos\phi + \tan\lambda\sin\phi)\right]^{2}}{\left[\sin\xi\sin\phi + C^2\cos\xi(\cos\phi + \tan\lambda\sin\phi)\right]^{2}}},
\end{equation}
\begin{equation}
  \bar{\nu} = \nu - \phi.
\end{equation}
\end{subequations}
\end{document}